\documentclass[ a4paper, revtex4-1,  nofootinbib, 10pt, twocolumn, floatfix, groupedaddress, aps, longbibliography, superscriptaddress] {quantumarticle}

\pdfoutput=1
\usepackage{times, mathrsfs, amsmath, amsfonts, graphics, graphicx, cancel, color, amsthm, bbm, mathtools, amssymb, physics}
\usepackage[nice]{nicefrac}
\usepackage[english]{babel}
\usepackage[margin=.75in]{geometry}
\usepackage[T1]{fontenc} 
\usepackage{capt-of}
\usepackage{xfrac,relsize,float}
\usepackage{comment}
\usepackage[normalem]{ulem}

\usepackage{appendix}

\usepackage[unicode=true, breaklinks=false, pdfborder={0 0 1}, backref=false, colorlinks=true, allcolors = blue, linkcolor=blue, citecolor=blue]{hyperref}

\def\bra#1{\mathinner{\langle{#1}|}}
\def\ket#1{\mathinner{|{#1}\rangle}}
\def\braket#1{\mathinner{\langle{#1}\rangle}}

\def\BraVert{\egroup\,\mid\,\bgroup}

\def\ketbra#1#2{|#1\rangle \!\langle#2|}

\newcommand{\lucacom}[1]{{\color{Red}{[Luca: #1]}}}

\definecolor{Blue}{rgb}{0,0,1}
\definecolor{Red}{rgb}{1,0,0}
\definecolor{Green}{rgb}{0,1,0}
\definecolor{darkgreen}{rgb}{0,.7,0}
\definecolor{Purp}{rgb}{.2,0,.2}
\definecolor{white}{rgb}{1,1,1}

\def\ecr{\textcolor{blue}}
\def\noteecr{\textcolor{darkgreen}}

\newcommand{\V}{\mathcal{V}}

\newcommand{\e}{{\scriptscriptstyle{E}}}

\usepackage{bbold}

\begin{document}
\title{Quantum Reference Frames for Lorentz Symmetry }

\author{Luca Apadula}\email[Luca Apadula: ]{  luca.apadula@univie.ac.at} 
\affiliation{Faculty of Physics, University of Vienna, Boltzmanngasse 5, 1090 Vienna, Austria }\affiliation{Institute for Quantum Optics and Quantum Information (IQOQI-Vienna), Austrian Academy of Sciences, Boltzmanngasse 3, 1090 Vienna, Austria}
\author{Esteban Castro-Ruiz}\email[Esteban Castro-Ruiz: ]{ 
ecastro@phys.ethz.ch }\affiliation{Institute for Theoretical Physics, ETH Zurich, Switzerland}
\author{$\check{\text{C}}$aslav Brukner}\email[Časlav Brukner: ]{ 
caslav.brukner@univie.ac.at }
\affiliation{Faculty of Physics, University of Vienna, Boltzmanngasse 5, 1090 Vienna, Austria }\affiliation{Institute for Quantum Optics and Quantum Information (IQOQI-Vienna), Austrian Academy of Sciences, Boltzmanngasse 3, 1090 Vienna, Austria}

\begin{abstract}

Since their first introduction, Quantum Reference Frame (QRF) transformations have been extensively discussed, generalising the covariance of physical laws to the quantum domain. Despite important progress, a formulation of QRF transformations for Lorentz symmetry is still lacking. The present work aims to fill this gap. We first introduce a reformulation of relativistic quantum mechanics independent of any notion of preferred temporal slicing. Based on this, we define transformations that switch between the perspectives of different relativistic QRFs. We introduce a notion of ``quantum Lorentz transformations'' and ``superposition of Lorentz boosts'',  acting on the external degrees of freedom of a quantum particle. We analyse two effects, superposition of time dilations and superposition of length contractions, that arise only if the reference frames exhibit both relativistic and quantum-mechanical features. Finally, we discuss how the effects could be observed by measuring the wave-packet extensions from relativistic QRFs.

\end{abstract}
\maketitle
\section{Introduction}

The formalism of quantum reference frames (QRFs) has received significant attention in recent years, both from the quantum gravity and from the quantum information and quantum foundations communities~\cite{QRFaharonov,QRS,Superselection,RF.superselection.QI,QRFresource,QRFdeformedsymmetries,QRFcommunication,QRFangelo,QRFangelo.riberio,ChangingQRF,QRFspin1,QRFnoncompactgroup,Miyadera,Loveridge,QRFmassiveobjects,QRF,QRFspin,quantum.clocks,QRFrelational,QRFrelational1,QRFunruheffect,Qclocks,QuantumEP2,QRFmassesinsuperposition,QuantumEP,QRFrelational2,QRFrelational3,QRFrelational4,QRFrelational5,QRFsubsystems,spacetimeQRF,QuantumEP1,Second.quantized.U-D}. The general idea behind QRFs is to extend the notion of reference frame symmetry transformations to the quantum realm. These transformations can be interpreted either as a change of description relative to a quantum system~\cite{QRF,QRFspin,QRFrelational,QRFrelational1,QRFsubsystems}, or more abstractly as symmetry transformations between different choices of quantum coordinates~\cite{QuantumEP,QuantumEP1,QuantumEP2,spacetimeQRF,QRFspacetime}

Most of the concrete scenarios involving QRF transformations have been studied in the domain of non-relativistic physics, Newtonian and post-Newtonian gravity. In Ref.~\cite{QRF} a quantum extension of Galilean symmetry as well as 
the notion of covariance of physical laws under these QRF transformations have been introduced.  Several works have reported  extensions  of QRFs to relativistic systems~\cite{QRFspin,spacetimeQRF,QRFrelational4}.  Ref.~\cite{QRFspin} introduces a ``quantum Wigner rotation'' that allows moving to the rest frame of a particle even if it is moving in a superposition of relativistic velocities with respect to the laboratory frame. This was used as a tool to solve the problem of an operational definition of spin in the relativistic regime. However, it focused only on the transformation of the internal degrees of freedom. Ref.~\cite{spacetimeQRF} introduces QRF transformations for spacetime translation symmetry and applies them to describe a quantum superposition of
special-relativistic time dilation to second order in $(p/mc)^2$.
Despite recent progress, a relativistic extension of QRF transformations, in the sense of Lorentz symmetry, is still missing.


In the present work we extend the quantum reference frame formalism to relativistic quantum systems carrying Lorentz symmetry. This is challenging because Lorentz transformations mix space and time, which calls for a framework that treats both space and time on the same footing.  Hence, we  use a spacetime representation of states with no preferred temporal slicing, which is inspired by a covariant formulation of quantum mechanics~\cite{CQM,CQM1}.
Using a ``coherent twirling'' approach~\cite{QRFrelational,QRFrelational1,QRFrelational2,QRFrelational3,RQM}, which  has been thoroughly discussed for unimodular Lie groups in ref.~\cite{de_la_Hamette_2020} as for locally compact ones in ref.~\cite{carette2023operational}, we define maps that transform between different QRFs for Lorentz symmetry. 
These maps can be understood as ``quantum Lorentz transformations'', giving rise to novel phenomena, such as superpositions of special relativistic time dilations and length contractions. Under a passive view, our symmetry transformations lead to a  definition of states on superposition of spacetime slices, which manifestly cannot be recovered within the standard Lorentz transformations, and resonates with recent extensions of the quantum framework to superpositions of semiclassical spacetime backgrounds~\cite{QRFmassesinsuperposition, QRFspacetime}. 
Our  results are obtained without resorting to any sub-relativistic (low speed) approximation, thereby, exactly complying
with the full  Lorentz symmetry. Other work~\cite{Smith_quantumtime, Smith_quantumtime1} has succeeded in providing quantum corrections for mass and proper time measurements, which arise from the quantum nature of the system under consideration. These proposals involve a detailed formulation of the POVM, rather than an extension of Lorentz symmetry to the quantum domain. 
As in ref.~\cite{Second.quantized.U-D}, where the authors investigated the decay of an excited particle in a superposition of relativistic velocities. Using the concept of Lorentz boosts for QRFs introduced in~\cite{QRFspin}, they transformed the corresponding POVM, initially defined in the particle's rest frame, back into the laboratory frame. This approach allowed them to observe effects ascribable to the quantum superposition of the probed state at different times.
Instead, in the following paper, a formulation of Lorentz symmetry is provided for our prescription of relativistic QRFs (RQRFs), thus making it possible to explore how spatial coordinates, e.g., clock and rod readings, as well as any other possible observable, transform between RQRFs, or in other words between quantum Lorentz observers.
We show that when the spacetime states are localised to geometrical points, these phenomena can be easily obtained from quantum-controlled (i.e. superpositions of) Lorentz coordinate transformations. Finally, the spacetime interval between  two arbitrary events is shown to be invariant under  quantum Lorentz transformations, extending the well-known result  of Minkowskian geometry to the quantum domain.

\section{Spacetime states and Probability} \label{Spacetime states and Probability}

We formulate relativistic quantum mechanics in a way that treats space and time symmetrically, in the spirit of the covariant formulation of quantum mechanics~\cite{CQM,CQM1}. For a recent alternative formulation, based on events rather than particles, see~\cite{Quantumevents}.
For simplicity, we consider $1+1$-dimensional spacetime (the $3+1$ case would require to treat the quantum reference frame for rotations additionally~\cite{QRFspinrotations}). 
We start from the spatial momentum eigenstates of a relativistic quantum particle with relativistic normalization, $\braket{p'|p}:=2E(p)\delta(p'-p)$, 
where $E(p) = \sqrt{p^2 + m^2} =: p^0$ is the energy (in units where $\hbar = c = 1$). In this basis, the resolution of identity reads
    $\mathsf{I}=\int \frac{dp}{2E(p)}\ketbra{p}{p}$,
where the integral sign with unspecified 
limits denotes integration over the real 
line.
Thus we can write the relativistic time evolution operator as
\begin{equation}\label{evo_operator}
\hat{U}(t):=e^{-i\hat{H}t}=\int \frac{dp}{2E(p)} e^{-iE(p)t}\ketbra{p}{p}.
\end{equation}
Consider now the relativistic 
propagator, restricted to positive 
energy only. It can be written as the
coordinate representation of Eq.~\eqref{evo_operator}
\begin{align}
    W(t',x';t,x)&:=
    \braket{x'|U(t'-t)|x}\label{propagator}\\
    &=:\braket{t',x'|t,x}, \label{innerproduct}
\end{align}
where we defined
\begin{align}\label{spacetimestate}
\ket{t,x}:=\hat{U}^\dagger(t)\ket{x}=\int \frac{dp}{2E(p)}e^{+iE(p)t-ipx}\ket{p}.
\end{align}
From Eq.~\eqref{innerproduct} we see that the inner product of kets in 
Eq.~\eqref{spacetimestate} results in  the 
relativistic propagator.

We now use linearity to define \textit{a quantum spacetime state}
\begin{equation}\label{f_state}
\ket{f}=\int dt dx f(t,x) \ket{t,x},
\end{equation}
where $f:\mathbb{R}^2\rightarrow\mathbb{C}$ defines 
the state localization through its support in spacetime.
We are now in the position to define the
notion of  \textit{relativistic wave function}:
\begin{align}
    \braket{t',x'|f}
   &=\int dtdx\;W(t',x';t,x)f(t,x)=:\psi_f(t',x') \label{rwf}
\end{align}
as the coordinate representation of the state in the Schrödinger picture
\begin{equation}\label{std_state}
 \ket{\psi_f(t')}=\int dx'\psi_f(t',x')\ket{x'}.
 \end{equation}
Its dynamics is described by the positive 
square root of Klein-Gordon equation
\begin{equation}
i\partial_{t'}\ket{\psi_f(t')}=\sqrt{\hat{p}^2+m^2}\ket{\psi_{f}(t')}. 
\end{equation}
 From Eqs.~\eqref{rwf} and \eqref{std_state} we may interpret
$f(t,x)$ as an extension of the set of initial conditions for a
quantum state to arbitrary spacetime regions. In particular, it enables to describe state preparations
that are not sharp in time. Therefore, $\ket{f}$ is not exclusively defined on any preferred spatial slicing, but rather within the spacetime volume of $f$'s support. 

We conclude this section showing that 
inner product of spacetime states is nothing but the standard Klein-Gordon inner product 
at an arbitrary time $t_0$ (see, e.g. Ref.~\cite{weinberg_1995}): 
\begin{align}
\braket{f'|f}&=\int dt'dx'\int dt dxf'^*(t',x') W(t',x';t,x)f(t,x)\label{scalarproduct}\\
&=i\int dx(\psi_{f'}^*(t_0,x)\partial_{t_0}\psi_f(t_0,x)-\textit{h.c.})
\label{scalarproduct}\\
&=\braket{\psi_{f'}(t_0)|\psi_f(t_0)}\quad\forall t_0\in\mathbb{R}.
\end{align}
Eq~\eqref{scalarproduct} shows that the inner product of 
spacetime functions on $\mathbb R^2$ reduces to 
the Klein-Gordon product defined on an
arbitrary spatial slicing, here denoted by the time label
$t_0$.

Upon the additional requirement of normalizability  
for the states in Eq.~\eqref{std_state}, we 
characterise set of allowed spacetime
functions $f$ as
\begin{equation}
\mathcal{E}=\{f: \mathbb{R}^2\rightarrow\mathbb{C}\,\, \text{such that}\,\,\psi_f(t,x)\in L^2(\mathbb{R}) \},
\end{equation}
namely, physically admissible spacetime functions are those for which $\psi_f(t_0,x):=\braket{t_0,x|f}$ is
normalizable with respect to the Klein-Gordon scalar product.
The set $\mathcal{E}$, endowed of the inner product~\eqref{scalarproduct}, is dense in the Hilbert 
space of states satisfying  the positive 
square root of Klein-Gordon equation. Hence, the  
normalized quantum state corresponding to $f\in \mathcal{E}$ is written as
\begin{equation}\label{normalised_state}
\ket{f_n}=\frac{\int dtdtf(t,x)\ket{t,x}}{\braket{f|f}^{\frac{1}{2}}}.
\end{equation}

\subsection{Probability}
Next we consider a complete 
observation test 
$\{\mathsf{P}_k\}_{k\in\mathcal{I}}$, where  each  $\mathsf{P}_k$ identifies a
POVM element, labelled by the corresponding
outcome $k$ contained in the set $\mathcal{I}$ of possible outcomes. We define the probability of outcome
 $k$ occurring given the spacetime state $\ket{f_n}$ as
\begin{align}\label{probability}
p(k|f_n)=\bra{f_n}\mathsf{P}_k\ket{f_n},
\end{align}
such that $\forall  k\in \mathcal{I}$ and 
$\forall f_n\in \mathcal{E}$ we have $p(k|f_n)\le 1$.
The completeness for any observation test 
$\sum_{k\in\mathcal{I}}\mathsf{P}_k=\mathbb{I}$
is ensured by the resolution of identity. That is
\begin{align}\label{complete_test}
\sum_{k\in\mathcal{I}} \bra{f_n}\mathsf{P}_k\ket{f_n}&=\bra{f_n}  \mathbb{I}\ket{f_n}=
\braket{f_n|f_n}=1.
\end{align}
Eqs.~\eqref{probability} and~\eqref{complete_test} show that probabilities
are well defined.

A straightforward example can be obtained by  setting $k=p$, so that
\begin{align}
    \mathsf{P}_p=\frac{\ketbra{p}{p}}{2E(p)},
\end{align}
having then
\begin{align}
    \int dp\, \mathsf{P}_p=\mathbb{I}.
\end{align}
\begin{figure*}[ht]
\centering
\includegraphics[width=340pt]{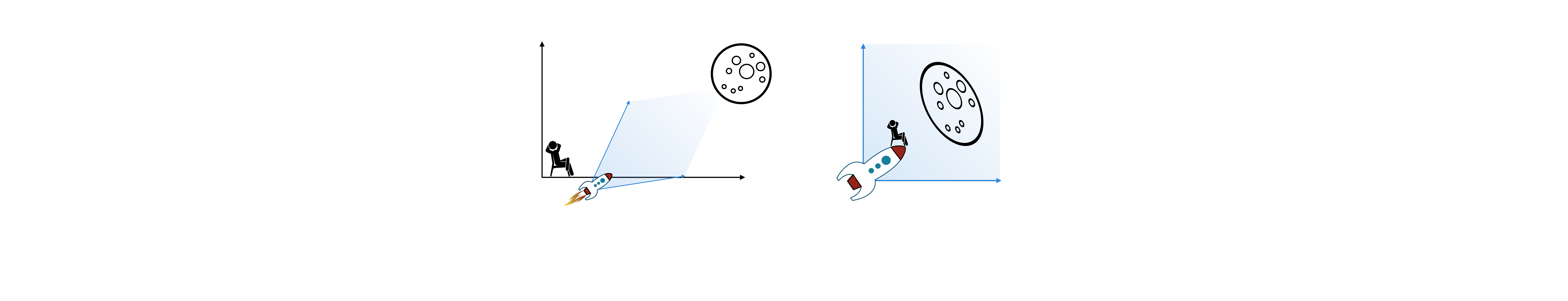}
\caption {\textit{Relativistic boost of a moon-shaped spacetime region}: (Left) A support of the spacetime state in form of the moon-shaped region as observed from the point of view of the observer ``sitting on the ground''. (Right) The same moon-shaped region as described from the point of view of the
observer ``sitting on top of the spaceship'' moving with a constant relative velocity with respect to the first observer.}
\label{stboosted}
\end{figure*}
As another example, consider a function $h\in\mathcal{E}$.  We can construct a 
complete test, by first defining a POVM 
element as
\begin{equation}\label{POVM}
    \mathsf{P}_h:=\ketbra{h_n}{h_n},
\end{equation}
corresponding to the detection of the 
system in the spacetime region identified
by the support of $h(t,x)$, with 
$\ket{h_n}$ defined 
as~\eqref{normalised_state}. 
The corresponding probability is then
\begin{align}
    &p(h|f)=\bra{f_n}\mathsf{P_h}\ket{f_n}=|\braket{h|f_n}|^2\label{spacetimeprobability}\\
    &= \frac{|\int dt'dx'\int dt dxh^*(t',x')\bra{x'}\hat{U}(t'-t)\ket{x}f(t,x)|^2}{\braket{h|h}^{\frac{1}{2}}\braket{f|f}^{\frac{1}{2}}}. \nonumber
\end{align}
According to the Cauchy-Schwarz
inequality, we have $p(h|f)\leq 1$ $\forall h,f \in \mathcal{E}$.
The complementary POVM element can be written as
\begin{align}
    \mathsf{P_{\Bar{h}}}:=\mathsf{I}-\mathsf{P}_h,
\end{align}
so that the completion to the identity follows directly.
As we will see in the next section, the scalar product is Lorentz invariant (see Eq.~\eqref{inv_scalarproduct}), and hence the invariance of the probabilities is
straightforwardly understood.
Despite being Lorentz invariant, unfortunately the probability formula in Eq.~\eqref{spacetimeprobability} suffers form the well-known problem of relativistic quantum theories based on particles~\cite{CQM1,Sorkin,Malament,Busch,Wightman}. Namely, the probability $p(h \vert f)$ for a particle to propagate from a spacetime region corresponding to $f$ to a space time region corresponding to $h$ does not vanish for spacelike separated regions. To resolve this problem, we need to extend our framework to quantum field theory. In this paper we focus on formulating Lorentz symmetry transformations for quantum reference frames, and leave the field-theoretic extension for future work.
\subsection{State transformation under Lorentz boost}

In this section we explore the transformation of spacetime states under the Lorentz group in 1+1 dimensions -- the one parameter group of relativistic boosts.
The descriptions of two inertial observers who move with speed $v$ relative to each other are related by
the relativistic boost $\Lambda_\alpha$,
specified by the rapidity $\alpha=\tanh^{-1}(v)$
as
\begin{equation}\label{matrix}
    \Lambda_\alpha=
    \begin{pmatrix}
    \cosh\alpha&-\sinh\alpha\\
    -\sinh\alpha&\cosh\alpha
    \end{pmatrix}.
\end{equation}
We denote by $U(\Lambda_\alpha)$ the unitary representation of the boost. Letting $U(\Lambda_\alpha)$ act
on the spacetime state~\eqref{f_state} we obtain
\begin{align}
U(\Lambda_\alpha)\ket{f}&=\int dt dx \;U(\Lambda_\alpha)\ket{t,x} f(t,x) \nonumber\\
&=\int dt dx \ket{\Lambda_{\alpha}(t,x)^\top} f(t,x)\\
&=\int d\tilde{t} d\tilde{x} \ket{\tilde{t},\tilde{x}} f_\alpha(\tilde{t},\tilde{x}) =: \ket{f_\alpha}, \nonumber
\end{align}
where we used the invariance of the volume element $dt dx$ and defined
\begin{align}
  (\tilde{t},\tilde{x})&:=\Lambda_{\alpha}(t,x)^\top:=\begin{pmatrix}
    \cosh\alpha&-\sinh\alpha\\
    -\sinh\alpha&\cosh\alpha
    \end{pmatrix}\begin{pmatrix}
    t\\
    x
    \end{pmatrix},\nonumber\\
      f_\alpha(\tilde{t},\tilde{x})&:=\left[f\circ\Lambda_{-\alpha}\right](\tilde{t},\tilde{x}).
\end{align}
To simplify  notation, we henceforth  omit the symbol 
of transposition, simply writing $\Lambda_{\alpha}(t,x)$.
A pictorial representation of a boosted spacetime  state is shown in  Fig.~\ref{stboosted}.

Consider a state defined on a spatial slice $\Sigma_t$, identified by a function of the form $f(t',x)=\delta(t-t')\phi(x)$,
\begin{equation}
\ket{f}=\int dt'dx  \ket{t',x} \delta(t-t')\phi(x).
\end{equation}
The corresponding  Lorentz transformed state is 
\begin{equation}\label{slicestate}
U(\Lambda_\alpha)\ket{f}=\int d\tilde{t}'d\tilde{x}\,\,\delta(t-\Lambda^0_{-\alpha}(\tilde{t}',\tilde{x}))\ket{\tilde{t}',\tilde{x}} \phi_\alpha(\tilde{x}),
\end{equation}
where $\phi_\alpha(\tilde{t}',\tilde{x}):=[\phi\circ\Lambda^1_{-\alpha}](\tilde{t}',\tilde{x})$ and  
$\Lambda^{0}_{\alpha}(t,x)$, $\Lambda^{1}_{\alpha}(t,x)$ stand for the time and 
space component of the Lorentz boosted coordinates, respectively.
Since $t'=\Lambda^0_{-\alpha}(\tilde{t}',\tilde{x})=\cosh\alpha\tilde{t}'+\sinh\alpha\tilde{x}$, we obtain
\begin{equation}\label{slicestate1}
U(\Lambda_\alpha)\ket{f}=\int \frac{d\tilde{x}}{\cosh\alpha}\ket{\tilde{t}'(\alpha,t,\tilde{x}),\tilde{x}}\phi_\alpha(\tilde{x}),
\end{equation}
where
\begin{figure*}[ht]
\centering
\includegraphics[width=380pt]{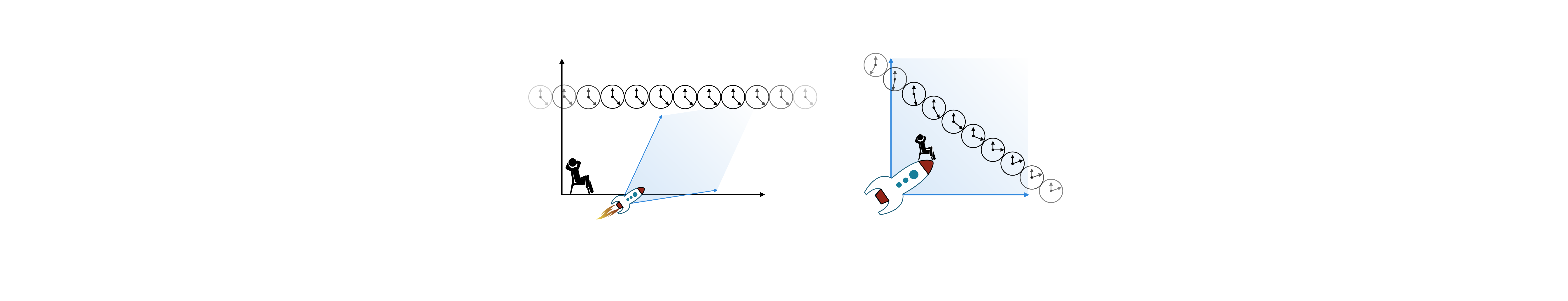}
\caption{\textit{Relativistic boost of a simultaneity surface}: (Left) The observer sitting at the ground, denoted as $C$ in the text, describes a state given on the simultaneity surface as illustrated with a plane of synchronised clocks. (Right) 
The observer on the top of the spaceship, labelled by $A$, moving with a relativistic speed, describes the state lying on a tilted spacetime hypersurface. 
}
\label{bslice}
\end{figure*}
\begin{align*}
   &\tilde{t}'(\alpha,t,\tilde{x})=(\cosh^{-1}\alpha t-\tanh\alpha\tilde{x}),\\
   &d\tilde{t}'d\tilde{x}\,\,\delta(t-\Lambda_{-\alpha}^0(\tilde{t}',\tilde{x}))=\frac{d\tilde{x}d\tilde t'}{\cosh\alpha}\delta(\tilde t' -\tilde{t}'(\alpha,t,\tilde{x})).
\end{align*}  
The last expression corresponds to the volume 
element of the transformed hypersurface
$\Lambda_\alpha(\Sigma_t)$. 
A graphical representation of the spacetime state both
before and after the Lorentz boost is given in 
Fig.~\ref{bslice}.

We conclude this section by showing that the invariance of the scalar product under Lorentz boost follows directly from the  relativistic propagator~\eqref{propagator}, which is 
manifestly Lorentz invariant.
\begin{align}\label{inv_scalarproduct}
&\braket{f_{\alpha}|f'_{\alpha}} \nonumber \\
&=\int dt dx\int dt' dx' f^*_{\alpha}(t',x')\bra{x'}U(t'-t)\ket{x}f'_{\alpha}(t,x)\nonumber \\
&=\int d\tilde{t} d\tilde{x}\int d\tilde{t}' d\tilde{x}' f^*(\tilde{t}',\tilde{x}')\bra{\tilde{x}'}U(\tilde{t}'-\tilde{t})\ket{\tilde{x}}f'(\tilde{t},\tilde{x}) \nonumber \\
&=\braket{f|f'}.
\end{align}
This implies that all the inertial observes agree on the transition amplitudes, as given by the scalar product of the states, and their normalization.

\section{Relativistic Quantum Reference Frame Transformations}\label{RQRF_formalism}

Next we  derive the transformation rules 
between RQRFs. For simplicity 
let us consider three 
physical systems, two of which serve as RQRFs
and the remaning one as a probe 
system.
We start with a generic state
\begin{align}\label{ext_state}
    \ket{\psi}_{\text{ext}}=\int d\alpha d\beta d\gamma \ket{\Lambda^1_\alpha k_A}\ket{\Lambda^1_\beta k_B}\ket{\Lambda^1_\gamma k_C}\psi(\alpha,\beta,\gamma).
\end{align}
Since we assume that there is no external reference frame for global Lorentz transformations, the state~(\ref{ext_state}) contains redundant information. The redundancy is encoded in the degrees of freedom that transform under the global action of the Lorentz group. We refer to these degrees of freedom as gauge. We will remove the redundancy via a group averaging technique~\cite{QRFrelational,QRFrelational1,QRFrelational2,QRFrelational3,RQM,spacetimeQRF}.
Operationally, the lack of an external reference frame leads to the invariance of the density martix $\rho$ under the action of the group, instead of the more stringent invariance of the state vector. In  the former case, one uses the incoherent $G$-twirl instead of the standard group averaging, and obtains more general QRF transformations~\cite{QRFsubsystems}. For the purpose of this work, however, we impose invariance of the state vector. It consists 
in projecting the state space into the subspace invariant under the action of the group, which in the present case is the 1+1 dimensional Lorentz group.
The group-averaged state \eqref{ext_state} is defined as
\begin{align}\label{perspective-less}
\mathcal{G}_{\text{Lor}}\left(\ket{\psi}_{\text{ext}}\right)
&:=\int d\omega \underset{i\in\{A,B,C\}}{\bigotimes}U^i(\Lambda_\omega)\ket{\psi}_{\text{ext}}:=\ket{\psi}_{\text{rel}}.\nonumber \\
\end{align}
We refer to the expression~\eqref{perspective-less}
as  ``perspective-neutral'', or relational state~\cite{QRFrelational,QRFrelational1,QRFrelational2,QRFrelational3,QRFrelational4,QRFrelational5} since it is  invariant under arbitrary global Lorentz boosts, that is, 
\begin{align}\label{invariant}
    \underset{i\in\{A,B,C\}}{\bigotimes}U^i(\Lambda_\beta)\ket{\psi}_{\text{rel}}
    =&\ket{\psi}_{\text{rel}},
\end{align}
as can be checked by direct calculation. Condition~\eqref{invariant} can be 
rephrased as 
$(\hat{K}_A+\hat{K}_B+\hat{K}_C)\ket{\psi}_{\text{rel}}=0$, where $\hat{K}_i=t_i\hat{p}^0_i+x_i\hat{p}^1_i$, $i\in\{A,B,C\}$ correspond to 
the generators of relativistic boosts.
Accordingly, 
the averaging procedure, showed in Eq.~\eqref{perspective-less}, projects
into the kernel of the  
global boost generator (center of global momentum).
This procedure leads to a relational description of the covariant degrees of freedom,  independent of any external Lorentzian observer. Consequently the only viable frames of reference are internal physical systems, and the remaining degrees of freedom are the relational ones.
Strictly speaking, our group-averaged states~\eqref{perspective-less} do not form a subspace of the Hilbert space and the averaging operator is not a projector. This is because the Lorentz group is not compact. Standardly, this problem is solved by defining the group averaging operator as a map to a different Hilbert space, i.e. the perspective-neutral state space, with a suitably defined inner product (see, for example Ref.~\cite{QRFrelational5}), which in the current case corresponds to
\begin{align}
    \braket{\psi|\psi'}_{\text{rel}}:&=\bra{\psi}\int d\omega \underset{i\in\{A,B,C\}}{\bigotimes}U^i(\Lambda_\omega)\ket{\psi'}_{\text{ext}}\\
    &=\bra{\psi}\delta(\hat{K}_A+\hat{K}_B+\hat{K}_C))\ket{\psi'}_{\text{ext}}
\end{align}
where $\delta(\hat{K}_A+\hat{K}_B+\hat{K}_C)$ constrains the states to the kernel of $\hat{K}_A+\hat{K}_B+\hat{K}_C$.
\subsection{Internal perspectives and quantum reference frame transformations}

Our goal is to explore how physics looks  
from  the perspective of an internal RQRF. 
Formally, we are looking for a 
definition of ``QRF perspective'', 
which describes the physics of all 
systems external to the QRF in 
question, and a transformation law, 
which we can use to change between 
different QRF perspectives. After 
their introduction in Ref.~\cite{QRF}, QRF 
transformations have been studied 
through the perspective neutral approach~\cite{QRFrelational,QRFrelational1,QRFrelational2,QRFrelational3,QRFrelational4,QRFrelational5}.
In the present case, we adapt the formalism of Ref.~\cite{QRFrelational} to ``jump'' from $\ket{\psi}_{\text{rel}}$ into the perspective of a given RQRF, say $C$.  To this end, we apply  a Lorentz boost controlled by the momentum of the chosen RQRF.
More concretely, the operator that maps $\ket{\psi}_{\text{rel}}$ to $C$'s perspective is 
\begin{align}\label{map_V}
\hat{\V}_C&=\int \frac{dp_C^1}{2E(p_C^1)} \ketbra{p_C^1}{p_C^1}\otimes {U^B}^\dagger(\Lambda_{\beta(p_C)})\otimes {U^A}^\dagger(\Lambda_{\beta(p_C)})  \\
&=\int \frac{dp_C^1}{2E(p_C^1)}  \ketbra{p_C^1}{p_C^1}\otimes {U^B}(\Lambda_{-\beta(p_C)})\otimes {U^A}(\Lambda_{-\beta(p_C)})\nonumber\\
&=\int d\alpha\ketbra{\Lambda^1_{\alpha}k_C}{\Lambda^1_{\alpha}k_C}\otimes U^B(\Lambda_{-\alpha})\otimes U^A(\Lambda_{-\alpha}).\nonumber
\end{align}
Here the two-vector momentum is 
expressed as $p_C:=(p_C^0, p_C^1)$, 
where $p_C^0$ refers to the 
relativistic energy of the particle $C$, 
while $\beta(p_C):=\tanh^{-1}(\frac{p^1_C}{p^0_C})$ refers to the rapidity.
The presence of a minus sign in 
$U^\dagger(\Lambda_\beta)=U(\Lambda_{-\beta})$
stems from the fact that to move the origin of the
frame of reference on $C$,  the
remaining particles must undergo a boost 
with the opposite velocity to that of $C$. 
We can further simplify the notation
by expressing $p_C=\Lambda_\alpha(k_C)$,
where $k_C:=(m_C,0)$ is the energy-momentum of $C$ in the co-moving frame. Here, $d\alpha=\frac{dp^1_C}{2E(p^1_C)}$.

The state of particles $B$ and $A$ as seen from RQRF $C$ is then defined as (see appendix~\ref{B} for details)
\begin{align}\label{psiam}
\ket{\psi}^{(C)}:=\hat{\V}_C\ket{\psi}_{\text{rel}}
=\ket{\Omega}_C\otimes\ket{\psi}_{BA}.
\end{align}
Crucial is the factorization of the resulting  
state, whereby the system $C$, left in the state $\ket{\Omega}:=\int\frac{dp_C^1}{2E(p_C)}\ket{p_C}$, is a complete uniform superposition of
momenta, i.e. the Lorentz boost invariant state,  factorized out from the remaining
systems. Furthermore, we notice 
Consequently, it contains no information about the degrees of freedom relative to $C$ and hence can be discarded. Thus, the state $\ket{\psi}_{BA}$ contains all the information about systems $B$ and $A$ ``as seen'' by $C$. 

We can change the perspective to $A$'s reference frame as well. All it takes is to define
 the operator $\hat{\V}_A$, analogously to $\hat{\V}_C$. We have now all the
elements to define a RQRF transformation from $C$ to $A$. The key 
observation is that $\hat{\V}_C$ is invertible, so we can start from $C$'s 
perspective, then transform to the  state of Eq.~\eqref{perspective-less}, and finally take the perspective of $B$'s frame. This procedure leads to the map 
$\hat{\mathcal{S}}_{C\rightarrow A}:\mathcal{H}^B\otimes\mathcal{H}^A\mapsto\mathcal{H}^B\otimes\mathcal{H}^C$,  defined by
\begin{align}\label{Lab}
\hat{\mathcal{S}}_{C\rightarrow A}:&=\bra{0}_A\circ \hat{\V}_A\circ\hat{\V}^\dagger_C \circ \ket{\Omega}_C\\
&=\int d\alpha \ketbra{\Lambda^1_{-\alpha} k_C}{\Lambda^1_\alpha k_A}\otimes U_B(\Lambda_{-\alpha}),
\end{align}
where $\bra{0}_A$ is arbitrarily chosen.
In general, the expression~\eqref{Lab} represents a Lorentz frame transformation between two RQRFs related through quantum relativistic boosts.

To the best of our knowledge,  the relational 
approach~\cite{QRFrelational,QRFrelational1,QRFrelational2,QRFrelational3,QRFrelational4,QRFrelational5} has been applied to 
those cases in which the group averaging 
technique and the global evolution  for the systems $A$, $B$ and $C$ as seen from the ``external'' point of view commute with 
each other. In other words, the group of transformations is a symmetry of the dynamics of the global system. In the present case averaging with respect to the Lorentz group does not commute with the 
unitary evolution of free dynamics~\eqref{evo_operator}, hence transformation~\eqref{map_V} is not a symmetry of the dynamics. In order to circumvent this, we apply map~\eqref{Lab}
directly on a perspectival physical scenario,
always called  $C$, which is described within the framework of section~\ref{Spacetime states and Probability}, 
leading to a specific RQRF's perspective.
Therefore, it is only observer $C$ who describes the dynamics as two free particles. In particular, the ``external'' observer describes three interacting particles. Nevertheless, by construction, the dynamics according to this observer is invariant under boosts.  

We next consider specific physical situations that give rise to novel phenomenology.

\subsubsection{Superposition of boosts}\label{sub_sup_boosts}

We  consider  the following physical scenario 
described by observer $C$:  two free systems 
are prepared in state
\begin{align}\label{sup_boosts.psi_c}
    \ket{\psi}_{AB}&=\int dt_Adx_A\int dt_Bdx_B\ket{f^A(t_A,x_A)}\otimes\ket{f^B(t_B,x_B)}\nonumber\\
    &:=\ket{f^A}\otimes\ket{f^B},
\end{align}
whose time evolution is governed by $U^A(t_A)\otimes U^B(t_B)$, as defined by Eq.~\eqref{evo_operator}. Here $\ket{f(t,x)}:=\ket{t,x}f(t,x)$. Let us assume that $A$ has been prepared in a
superposition of two sharp values of momenta $p_i:=\Lambda_{\omega_i}k_A$ with $i=1,2$. For simplicity, we chose a spacetime 
function $f^A$ in Eq.~\eqref{sup_boosts.psi_c} such that its Fourier transform for variable $p^1$ is (ignoring normalisation) 
\begin{align}\label{sharp_momenta}
\tilde{f}^A(t,p^1)=g_A(t) E(p^1)(\delta(p^1-p^1_1)+\delta(p^1-p^1_2)),
\end{align}
for some function $g_A(t)$. Then we move to  $A$'s perspective 
applying transformation~\eqref{Lab} to the state~(\ref{sup_boosts.psi_c}), i.e.
\begin{align}
\ket{\psi}_{BC}:=\hat{\mathcal{S}}_{C\rightarrow A}\ket{\psi}_{AB}.
\label{karta}
\end{align}
\begin{figure*}[ht]
\centering
\includegraphics[width=340pt]{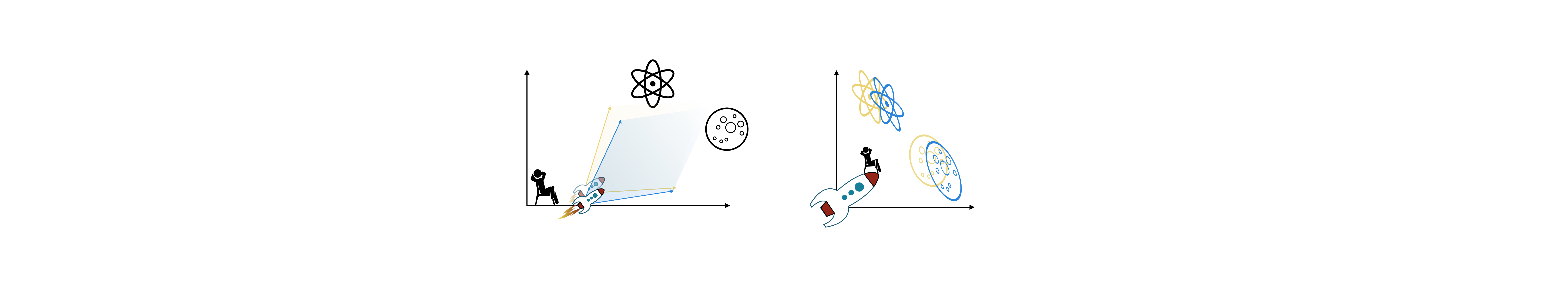}
\caption{\textit{Superposition of relativistic boosts} (Left)
The space-time supports of the states of two particles with the shapes of the moon and the atom in the laboratory reference frame. (Right) The two particles are entangled in the new reference frame of the spaceship moving in superposition of Lorentz velocities. The effect of the change of perspective becomes visible through (a) the change in shape of the supports of the spacetime states and (b) the entanglement between the two particles, which is illustrated by the supports correlated in colour.}
\label{sup_boost}
\end{figure*}
\begin{figure*}[ht]
\centering
\includegraphics[width=340pt]{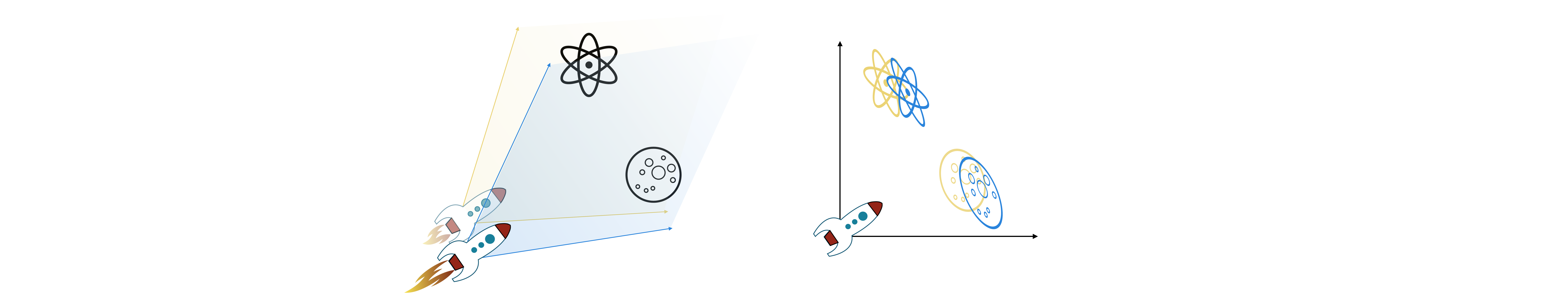}
\caption{\textit{Active and passive transformation of the state under a superposition of Lorentz boosts}: The transformed state can either be understood as given within a fixed spacetime support expressed in a superposition of two coordinates of two Lorentz frames (passive transformation, left) or as an entangled state in a pair of spacetime supports expressed in the coordinates of a single Lorentz frame (active transformation, right).}
\label{passive_active}
\end{figure*}
The state of $B$ and $C$ relative to $A$ 
has the following form (see  appendix~\ref{superposition of boosts} for a detailed derivation):
\begin{widetext}
\begin{align}
\ket{\psi}_{BC}&=\tilde{g}_C(\Lambda^0_{\omega_1}k_C)\ket{\Lambda^1_{-\omega_1}k_C}\otimes\ket{f^B_{-\omega_1}}+\tilde{g}_C(\Lambda^0_{\omega_2}k_C)\ket{\Lambda^1_{-\omega_2}k_C}\otimes\ket{f^B_{-\omega_2}}\label{superposition_boosts}\\
&=\int dt_B dx_B \left(\tilde{g}_C(\Lambda^0_{\omega_1}k_C)\ket{\Lambda^1_{-\omega_1}k_C}\otimes \ket{\Lambda^1_{-\omega_1}(t_B,x_B)}+\tilde{g}_C(\Lambda^0_{\omega_2}k_C)\ket{\Lambda^1_{-\omega_2}k_C}\otimes \ket{\Lambda^1_{-\omega_2}(t_B,x_B)}\right)f^B(t_B,x_B)\label{passivo}\\
&=\int d\tilde{t}_Bd\tilde{x}_B\ket{\tilde{t}_B,\tilde{x}_B}\otimes\left(\tilde{g}_C(\Lambda^0_{\omega_1}k_C)\ket{\Lambda^1_{-\omega_1}k_C}f^B(\Lambda_{\omega_1}(\tilde{t}_B,\tilde{x}_B))+\tilde{g}_C(\Lambda^0_{\omega_1}k_C)\ket{\Lambda^1_{-\omega_2}k_C}f^B(\Lambda_{\omega_2}(\tilde{t}_B,\tilde{x}_B))\right),\label{attivo}
\end{align}
\end{widetext}
where we defined $(\tilde{t},\tilde{x}):=\Lambda_{-\omega}(t,x)$, $t_C:=\frac{m_A}{m_C}t_A$
and $g_C(t_C):=\frac{m_C}{m_A}g_A(\frac{m_C}{m_A}t_A)$, 
whose Fourier transform is 
$\tilde{g}_{C}(\Lambda_{-\omega_i}k_C)$.
Note that  $\ket{\psi}_{BC}$ 
cannot be obtained from $\ket{\psi}_{AB}$ by means of 
a classical Lorentz boost. Indeed, 
particle $B$ undergoes a quantum
superposition of
two different boosts, labelled by $\omega_i$ with $i=1,2$, which are 
quantum-controlled
by the momenta of  RQRF $C$, as 
Eq.~\eqref{superposition_boosts} 
shows.  A pictorial 
representation of such scenario is 
displayed in Fig.~\ref{sup_boost}.
Eqs.~\eqref{passivo} 
and~\eqref{attivo} correspond, 
respectively, to the passive and active point of view of the coordinate 
transformation of $B$, as illustrated in 
Fig.~\ref{passive_active}.
\begin{figure*}[ht]
\centering
\includegraphics[width=400pt]{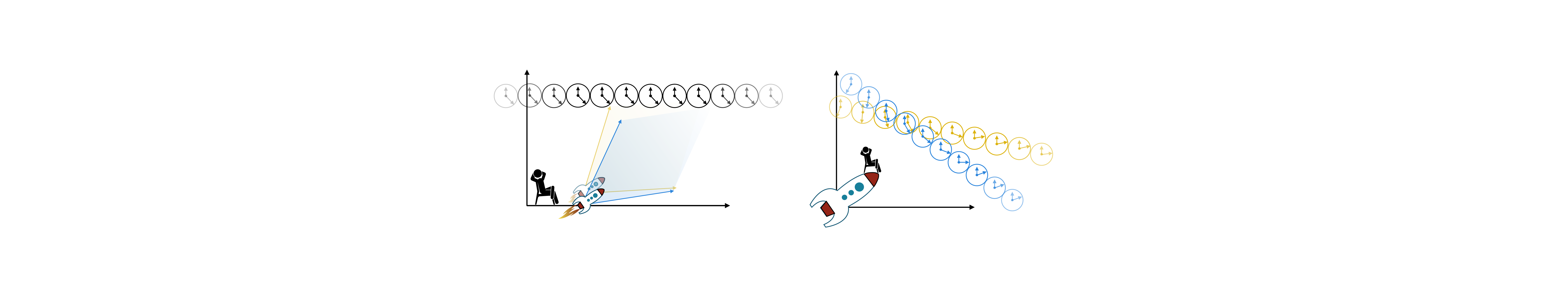}
\caption{\textit{Simultaneity surface in a superposition of relativistic boosts}: (Left) A plane of synchronised clocks and a space ship in a superposition of velocities are displayed in the reference frame of the observer $(C)$ on the ground . (Right) In the reference frame of the observer $(A)$ in the spaceship  the previous simultaneity surface transforms into in a superposition of tilted hypersurfaces, identified by the planes of blue and yellow clocks.}
\label{bes}
\end{figure*}
The passive transformation is depicted in 
left panel of the figure: two fixed 
spacetime volumes are described from the 
perspective of a QRF in a superposition of 
two different Minkowskian spacetime frames.
The right panel  shows, 
instead, the active point of view, 
where one keeps fixed a single spacetime 
frame and transforms actively the 
state to the one entangled in the two volumes.
The state in 
Eq.~\eqref{superposition_boosts} 
manifestly displays correlations in 
spacetime. Such correlations (i.e. in a single basis) can always be reproduced by a separable (classically correlated) state in space-time.
However, state~\eqref{superposition_boosts} additionally exhibits a correlation in the energy-momentum basis, which cannot be reproduced by the separable state. Indeed we have
\begin{widetext}
\begin{align}
    \ket{\psi}_{BC}
    &=\int d\beta \left(\tilde{g}_C(\Lambda^0_{\omega_1}k_C)\ket{\Lambda^1_{-\omega_1}k_C}\otimes\ket{\Lambda^1_{-\omega_1+\beta}k_B}+\tilde{g}_C(\Lambda^0_{-\omega_2}k_C)\ket{\Lambda_{-\omega_2}k_C}\otimes\ket{\Lambda^1_{-\omega_2+\beta}k_B}\right)\tilde{f}^B(\Lambda_{\beta}k_B),
\end{align}
\end{widetext}
and \mbox{$\tilde{f}^B(\Lambda_\beta k_B):=\int dt_Bdx_Be^{i\Lambda^0_\beta k_Bt_B+\Lambda^1_\beta k_Bx_B}f^B(t_B,x_B)$} is the Fourier transform of the spacetime
function $B$. The evolution operator, 
governing the dynamics of the composed 
system $BC$ relative to $A$'s
perspective, can be recovered from the 
time operator relative to $C$ as
\begin{align}
    &U^{(A)}:=\hat{\mathcal{S}}_{C\rightarrow A}U^A(t_A)\otimes U^B(t_B)\hat{\mathcal{S}}^\dagger_{C\rightarrow A}\\
    &=\int d\alpha\ketbra{\Lambda^1_{-\alpha}k_C}{\Lambda^1_{-\alpha}k_C}e^{i\Lambda^0_{\alpha}k_At_A}\otimes U^B(\Lambda_{-\alpha}(t_B,0)). \nonumber
\end{align}
We note that the time operator in QRF C is an entangling operator and therefore does not describe evolution of two free particles, although such an evolution was assumed from A's point of view. In other words, the free Hamiltonian is not invariant under quantum Lorentz boosts as introduced here. This is reminiscent of the situation in the theory of non-relativistic QRFs, where the free Hamiltonian is not invariant under  quantum space translations~\cite{QRF}.

\subsubsection{Superposition of spacetime slices}

In the Schr\"{o}dinger picture of standard quantum mechanics, a state is defined at a spacelike hypersurface. This hypersurface corresponds to a fixed time $t$ relative to an observer's reference frame, which we denote here by $C,$ and it is called $C$'s simultaneity surface. 
Consider now a second observer, with a reference frame $A$, moving with velocity $v$ with respect to $C$. According to special relativity,  $C$'s hypersurface does not correspond to a simultaneity surface for $A$. 
Instead, it is a tilted hypersurface along which both space and time coordinates change. Consider now the case in which $A$ moves in a superposition of
velocities. How does the description of physics change between QRFs? How does the quantum state and the resulting spacetime volume transform with the change of perspective from one observer to another? 

To answer these questions, we start from the perspective of $C$ and the state 
\begin{equation*}
    \ket{\psi_{AB}}=\ket{f^A}\otimes\ket{f^B},
\end{equation*}
where the support of $B$'s spacetime function is the simultaneity surface  $\Sigma_{t_B}$ (in $C$'s reference frame), defined by a spacetime function $f^B$
\begin{equation}\label{slice_b}
  \ket{f^B}=\int dt dx \, \ket{t,x} \delta(t-t_B)\phi^B(x),
\end{equation}
with $\phi(x)^B\in \mathcal{L}^2(\mathbb{R})$. 
The state of RQRF $A$ is assumed to be a superposition of around two sharp values of momenta,
similarly to the scenario in~\ref{sub_sup_boosts}. Hence, 
\begin{equation}\label{sharpsupt_A}
\ket{f^A}=\int dt dx \ket{t,x} \delta(t-t_A)\phi^A(x),
\end{equation}
where the Fourier transformed of the space function $\phi^A(x)$ satisfies the condition
Eq.~\eqref{sharp_momenta}. 
\begin{figure*}[ht]
\centering
\includegraphics[width=380pt]{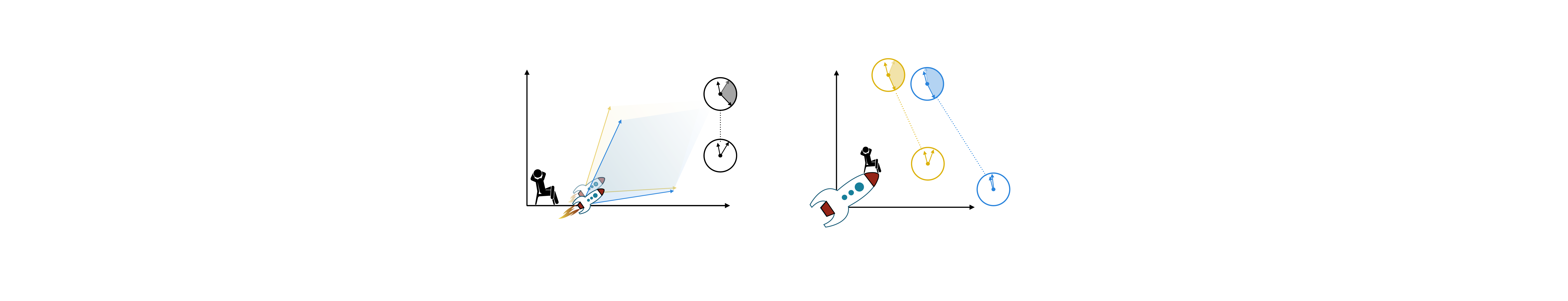}
\caption{\textit{Superposition of special-relativistic time dilation}: (Left) A time interval is given by the gray shadowed wedge of the clock for the observer ($C$) sitting at the ground. (Right) The observer  moving in the spaceship ($A$)  describes that time on the clock (i.e. the clock hand) to be in a superposition of dilated time intervals.}
\label{td}
\end{figure*}
We next move to $A$'s point of view acting with $\hat{\mathcal{S}}_{C\rightarrow A}$,
namely (see appendix~\ref{superposition of simultaneity surfaces} for derivation)
\begin{widetext}
\begin{align}\label{psibc_supsli1}
\ket{\psi_{BC}}&= \hat{\mathcal{S}}_{C\rightarrow A} \ket{\psi_{AB}}
=\ket{\Lambda^1_{-\omega_1}k_C}e^{i\Lambda^0_{\omega_1}k_Ct_C}\otimes\ket{f^B_{-\omega_1}}+\ket{\Lambda^1_{-\omega_2}k_C}e^{i\Lambda^0_{\omega_2}k_Ct_C}\otimes\ket{f^B_{-\omega_2}}
=\ket{C_1}\otimes\ket{f^B_{-\omega_1}}+\ket{C_2}\otimes\ket{f^B_{-\omega_2}},
\end{align}
\end{widetext}
where we denoted $\ket{C_i}:=\ket{\Lambda^1_{-\omega_i}k_C}e^{i\Lambda^0_{\omega_i}k_Ct_C}$ for  simplicity of notation.
We see that, relative to $A$, the state of particle $C$ lies in a spatial slice labelled by $t_C:=\frac{m_A}{m_C}t_A$ in both branches with momentum $\Lambda^1_{-\omega_i}k_C$, $i=1,2$.
The kets $\ket{f_{-\omega{i}}^B}$, $i=1,2$, are of the form of  Eqs.~\eqref{slicestate} and~\eqref{slicestate1}. From $A$'s reference frame,  $\ket{\psi_{BC}}$ describes an entangled state of $B$ and $C$, such that the state of $B$ is correlated with the state of sharp velocity of $C$. 
Most importantly, the state of $B$, 
previously corresponding to a single simultaneity 
surface of $C$, now lies on a superposition of tilted hypersurfaces relative RQRF $A$. For a pictorial representation of such state, we refer to Fig.~\ref{bes}.

In the next sections we  analyse some of the distinctive 
special-relativistic phenomena, such as dilation 
of time intervals and contraction of spatial 
lengths, in the case when RQRFs are in states of superposed momenta, using the formalism we have developed so far. 

\subsection{Superposition of special-relativistic time  dilations}
Special relativity predicts that for an observer 
in an inertial frame, a clock moving relative to 
her will tick slower than a clock at rest in her 
frame of reference. This is known as special-relativistic time dilation. We consider the
quantum generalisation of this phenomenon. 

In our framework, we identify an ``event'' with the outcome of measuring the space-time position of a particle or with a preparation of a well-localised spacetime state. For example, a measurement of the spacetime location of a particle by a POVM~\eqref{POVM} corresponding to the detection of the particle in a highly localised spacetime region corresponds to an event located at the spacetime point where the particle is found. To analyse time dilation effects we consider two events happening at the same point in space but at two different instances of time, according to  observer $C$. A state describing such pair of events is given by 
\begin{equation}
\ket{\psi}^{(C)}=\ket{f^{B_1}}\otimes\ket{f^{B_2}}\otimes\ket{f^A}, 
\end{equation}
with particles $B_1$ and $B_2$ defined by  the
following spacetime functions
\begin{align}
f^{B_1}(t,x)&\approx \delta(t-t_1)\delta(x-x_0),\\
f^{B_2}(t,x)&\approx \delta(t-t_2)\delta(x-x_0).
\end{align}
Hence, we identify the two events with the detections of particle $B_1$ at $(t_1,x_0)$ and particle $B_2$ at $(t_2,x_0)$. 
In addition, we assume that particle $A$ is again prepared in a superposition of two sharp values of momenta, as described by state~\eqref{sharp_momenta}.


We next move into the frame of reference of $A$ via the map 
$\hat{\mathcal{S}}_{C \rightarrow A}$, obtaining 
\begin{equation}\label{sup_timedilation}
\ket{\psi}^{(A)}=\ket{f_{-\omega_1}^{B_1}}\otimes\ket{f_{-\omega_1}^{B_2}}\otimes\ket{f^C_{-\omega_1}}+\ket{f_{-\omega_2}^{B_1}}\otimes\ket{f_{-\omega_2}^{B_2}}\otimes\ket{f^C_{-\omega_2}},
\end{equation}
where, for example 
\begin{equation}
    \ket{f^{B_1}_{-\omega_1}}=\int d\tilde{t}d\tilde{x} \, \ket{\tilde{t}_1,\tilde{x}}\delta(\Lambda_{\omega_1}^0(\tilde{t}_1,\tilde{x})-t_1)\delta(\Lambda_{\omega_1}^1(\tilde{t}_1,\tilde{x})-x_0)
\end{equation}
with $t=\Lambda_{\omega_1}^0(\tilde{t}_1,\tilde{x})$, $x=\Lambda_{\omega_1}^1(\tilde{t}_1,\tilde{x})$ and
\begin{align}
    &\delta(\Lambda_{\omega_1}^0(\tilde{t},\tilde{x})-t_1)=\frac{ \delta(\tilde{t}_1-\tanh{\omega_1}\tilde{x}-\cosh^{-1}{\omega_1}t_1)}{\cosh\omega_1} \label{deltauno}\\
    &\delta(\Lambda_{\omega_1}^1(\tilde{t}_1,\tilde{x})-x_0)=\frac{\delta(\tilde{x}-\tanh\omega_1\tilde{t}_1-\cosh^{-1}{\omega_1}x_0)}{\cosh\omega_1} \label{deltados}
\end{align}
and likewise for $\omega_2$. The same computations hold for $B_2$. Solving for $\tilde{t}_1$ and $\tilde{t}_2$ in Eqs.~\eqref{deltauno} and~\eqref{deltados}, 
\begin{figure*}[ht]
\includegraphics[width=500pt]{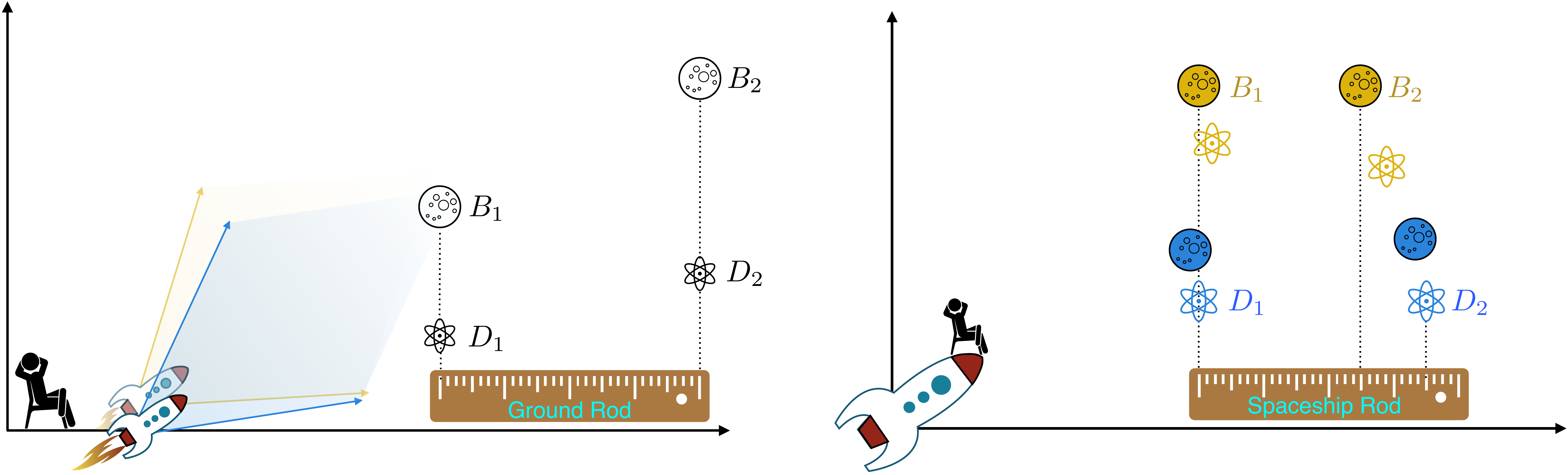}
\caption{\textit{Superposition of special-relativistic space contractions} (Left) The observer $C$ on the ground measures, with his rod, the length separating the moon-shaped and atom-shaped spacetime events. (Right) Upon jumping on the spaceship $A$, moving in a superposition of velocities, the observer uses the new ``spaceship rod'' for probing the 
space separation of the resulted superposition of pars of simultaneous events: in the yellow branch of the superposition they measures the spatial distance between the moon-shaped spacetime events, while in the blue one, the distance  between the atom-shaped spacetime events.}
\label{rod}
\end{figure*}
we find  $\tilde{t}_1=\cosh{\omega_1}(\tanh{\omega_1}x_0-t_1)$ for $B_1$ and 
$\tilde{t}_2=\cosh{\omega_1}(\tanh{\omega_1}x_0-t_2)$
for $B_2$.
Now we can compute the time intervals in the perspective of $A$, 
\begin{equation}\label{suptimeinterval}
    \Delta\tilde{t}_{i}=\tilde{t}_2-\tilde{t}_1=\cosh{\omega_{i}}(t_2-t_1)=\gamma(\omega_{i})\Delta t
\end{equation}
for each branch $i=1,2$ of the superposition, and with the time interval $\Delta t = t_2-t_1$ measured in $C$'s reference frame.

We conclude that, 
Eq.~\eqref{sup_timedilation} describes two particles located in spacetime such that their temporal separation is
in a superposition of two time intervals. The two intervals that result
from a special-relativistic dilation of a given time interval $\Delta t$ in the rest frame for two values of the Lorentz boost. A graphical illustration of this scenario is shown
in Fig.~\ref{td}.

\subsection{Superposition of special-relativistic length contractions}


We next consider a quantum
superposition of relativistic length contractions. Consider an observer $C$ and a rigid rod in a 
state of relative 
motion with respect to $C$. $C$ determines the rod's length $\Delta x_C$ by 
measuring the position of its ends at the same time, i.e.  
$\Delta t_C=0$. According to special relativity, a second observer $A$, who is co-moving with the rod, measures a length $\Delta x_A=\gamma(v)\Delta x_C$. 
In other words, the length measured by the moving observer, relative to the one measured by the observer at rest, is contracted:
$\Delta x_C=\frac{1}{\gamma(v)}\Delta x_A$. Using relativistic QRF transformations, we now extend this landmark  result to situations where observers are moving in a superposition of velocities with respect to each other. 

Let us consider two well localised particles 
prepared  in the reference frame of $C$,
such that  particle $B_1$ lies
at spacetime point $(t_{B_1}, x_{1})$, while  $B_2$ at $(t_{B_2}, x_{2})$. Their spatial
and temporal separations are $\Delta x=x_2-x_1$ and 
$\Delta t_B=t_{B_2}-t_{B_1}$, respectively. 
Again, a second pair of particles, $D_1$ and 
$D_2$, are prepared in the same space locations
at different times, namely 
$(t_{D_1},x_1)$ and $(t_{D_2},x_2)$ so that
the space separation is the same as for $B_1$ and $B_2$, while $\Delta t_D=t_{D_2}-t_{D_1}$. 
Finally, we include RQRF $A$, 
which is in a superposition of two sharp values 
of velocities $v_b = \frac{\Delta t_B}{\Delta x}$ and $v_d = \frac{\Delta t_D}{\Delta x}$ as 
in Eq.~\eqref{sharp_momenta}. 
The state as seen from RQRF $C$  is given by 
\begin{align}
    \ket{\psi}^{(C)}=\ket{f^{B_1}}\otimes\ket{f^{B_2}}\otimes\ket{f^{D_1}}\otimes\ket{f^{D_2}}\otimes\ket{f^A}.
\end{align}
We now move to $A$'s frame of reference 
by means of $\hat{\mathcal{S}}_{C\rightarrow A}$, obtaining
\begin{align}\label{suprod}
\ket{\psi}^{(A)}=\sum_{i=b,d}
\underset{j=1,2}{\bigotimes}\ket{f_{-\omega(v_i)}^{B_j}}\otimes\ket{f_{-\omega(v_i)}^{D_j}}\otimes\ket{f_{-\omega(v_i)}^{C}},
\end{align}
where the rapidity is expressed as a function of the velocity via the relation $\omega(v)=\tanh^{-1}(v)$.
The spacetime  states are illustrated in 
Fig.~\ref{rod}.
We chose the two pairs of events, $(B_1, B_2)$ and $(D_1, D_2)$, such that one pair of events lies on a simultaneity 
surface of $A$ in each branch of state~\eqref{suprod}. More precisely, the pair of events  $(B_1, B_2)$ lies on the simultaneity surface defined by the boost by $v_b$, and similarly the pair $(D_1, D_2)$ lies on the 
simultaneity surface defined by the boost by $v_d$. This is the case when  $\Delta t_B=v_b\Delta x$ and $\Delta t_D=v_d\Delta x$.

In the first branch of the superposition~(\ref{suprod}), the events represented by $f_{-\omega(v_b)}^{B_1}$ and $f_{-\omega(v_b)}^{B_2}$ are simultaneous, so that their spatial separation can be considered as the ``length of the rod''. 
The new time coordinates are given by 
$t_{B_1}'=\cosh(\omega(v_b))t_{B_1}-\sinh(\omega(v_b))x_1$, and 
$t_{B_2}'=\cosh(\omega(v_b))t_{B_2}-\sinh(\omega(v_b))x_2$
which we chose to be the same,    
$t'_{B_1}=t'_{B_2}$, with a suitable choice of 
$v_b$. Hence, one has $\Delta t_B=\tanh(\omega(v_b))\Delta x$ that, together 
with 
$x'_1=\cosh(\omega(v_B))x_1-\sinh(\omega(v_b))t_{B_1}$ and
$x'_2=\cosh(\omega(v_b))x_2-\sinh(\omega(v_B))t_{B_2}$,
lead to
$\Delta x'=x'_2-x'_1
    =\gamma(v_b)^{-1}\Delta x$
which is exactly the special-relativistic length
contraction of the $C$'s rod $(\Delta x)$, 
measured by $A$'s rod $(\Delta x')$. 

Now we do the same  for the second branch 
of~\eqref{suprod}, where ``the rod'' is identified
by two simultaneous events, $f^{D_1}_{-\omega(v_d)}$ and $f^{D_2}_{-\omega(v_b)}$ with the spacetime coordinates $(t'_{D_1},x'_{D_1})$ and 
$(t'_{D_2},x'_{D_2})$, respectively. Hence, we 
find the same relation
\begin{align}
    \Delta x'_D=x'_{D_2}-x'_{D_1}=\gamma(v_d)^{-1}\Delta x,
\end{align}
which, as before, represents the contraction of
$C$'s rod experienced by $A$. We therefore conclude that Eq.~\eqref{suprod} describes a 
superposition of different length contractions of
the same rod. In Appendix A we show that the same effect of superposition of length contractions is behind the phenomenon of superposition of wave-packet widths.

\subsection{Quantum relativistic coordinate transformations}

In this section we develop a quantum generalization 
of special relativistic coordinate  transformations, 
 where the inertial observers can be in a 
superposition of velocities. To this end, we first give a kinematical prescription of spacetime states, simply removing  the relativistic time evolution from  definition~\eqref{spacetimestate}, and then restrict 
them  only to have point-wise supports. They 
correspond to spacetime ``events'' as defined in this 
work. Let 
\begin{equation}\label{event}
\ket{\textit{event}}^C:=\ket{t}\otimes\ket{x}=\ket{(t,x)}_{\textit{E}}=:\ket{\Bar{x}}_{\textit{E}},
\end{equation}
be a ``coordinate state''  for  observer $C$, where  
$\ket{\Bar{x}}$ has no dynamical content, but only
represents the coordinates of the geometrical point 
in spacetime that label the event. The $(t,x)$ labels can be viewed as readings of quantum systems, concerned as clock and rod respectively, whose dynamical is ignored. 

Consider now a new laboratory frame, $A$,
moving in a superposition of velocities with respect 
to $C$:
\begin{equation}\label{lab in sup}
\ket{\textit{lab}_{\textit{A}}}^C=\ket{v^1}_\textit{A}+\ket{v^2}_\textit{A}.
\end{equation} 
While the event is identify by the readings of a quantum clock and rod, the reference frame, i.e. the laboratory, is rather identified by a state of velocity. This is related to the momentum of a quantum system via $p=v\gamma m$,
from which $v=\frac{p/m}{\sqrt{1+\frac{p^2}{m^2c^2}}}$. Eventually, the corresponding Hamiltonian is irrelevant for the following discussion.  
The joint state describing both the laboratory and two spacetime events, relative to $C$, is given by
\begin{align}
\ket{\psi}^C&=\ket{\textit{lab}_{\textit{A}}}^C\otimes\ket{\textit{event}_1}^C\otimes\ket{\textit{event}_2}^C\\
&=(\ket{v^1}_{\textit{A}}+\ket{v^2}_{\textit{A}})\ket{\Bar{x}_1}_{\textit{E}_1}\ket{\Bar{x}_2}_{\textit{E}_2}.\label{events for C}
\end{align}
The coordinate transformation that switches from the description of observer $\textit{C}$ to the that of $\textit{A}$, can be obtained by straightforwardly
adjusting the map in Eq.~\eqref{Lab} as follows
\begin{equation}\label{quantum-controlled Lorentz coordinate}
S_{\textit{C}\textit{A}}:=\mathcal{P}_{\textit{C}\textit{A}}\circ\hat{\Lambda}_{-\hat{v}_\textit{A}},
 \end{equation}
where $\mathcal{P}_{\textit{C},\textit{A}}$ acts on  $\ket{\textit{lab}_{\textit{A}}}^C$ as the parity-swap operator~\cite{QRF,QRFspin},
returning the state of laboratory $C$ with respect to $A$ as
\begin{equation}
\mathcal{P}_{\textit{C}\textit{A}}\ket{\textit{lab}_{\textit{A}}}^C=\ket{-v^1}_\textit{C}+\ket{-v^2}_\textit{C}=:\ket{\textit{lab}_{\textit{C}}}^{\textit{A}}.
\end{equation}
The transformation $\hat{\Lambda}_{-\hat{v}_\textit{A}}:=\int dv\ketbra{v}{v}_{\textit{A}}\otimes\hat{\Lambda}_{-v}$ is a quantum-controlled Lorentz coordinate transformation 
\begin{equation}
    \hat{\Lambda}_{-v}:=\int d\Bar{x} \ket{\Lambda_{-v}\Bar{x}}\bra{\Bar{x}},
\end{equation}
with $\Lambda_{-v}\Bar{x}=\Lambda_{-v}(t,x)=(t\cosh{\alpha(-v)}-x\sinh{\alpha(-v)},x\cosh{\alpha(-v)}t\sinh{\alpha(-v)})$. 
That is the action of $\Lambda_{-v}$ (see Eq.~\eqref{matrix}) on $\Bar{x}$
gives the coordinates in the new frame of reference.
The operator $\hat{\Lambda}_{-\hat{v}}$ transforms 
coherently the event's coordinate state $|\Bar{x}\rangle$ to $\ket{\Lambda_{-v}\Bar{x}}$, depending on the velocity of the laboratory. 

Let us now transform the state in Eq.~\eqref{events for C},
written in the QRF of $C$, to the QRF of $A$. Using 
Eq.~\eqref{quantum-controlled Lorentz coordinate}, we 
obtain
\begin{multline}
    S_{\textit{C}\textit{A}}\ket{\psi}^C=\ket{-v^1}_\textit{C}\ket{\Lambda_{-v^1}\Bar{x}_1}_{\textit{E}_1}\ket{\Lambda_{-v^1}\Bar{x}_2}_{\textit{E}_2}\\
    \quad+\ket{-v^2}_\textit{C}\ket{\Lambda_{-v^2}\Bar{x}_1}_{\textit{E}_1}\ket{\Lambda_{-v^2}\Bar{x}_2}_{\textit{E}_2}
    =\ket{\psi}^{\textit{A}}.
\end{multline}
Now we show that the spacetime distance between events defined in Eq.~\eqref{event}
is invariant under transformation~\eqref{quantum-controlled Lorentz coordinate}. Let us introduce the 
``spacetime distance'' operator 
\begin{align}\label{distance operator}
\hat{\mathcal{D}}^{C}:&=\mathbb{I}_{\textit{lab}_A}\otimes\hat{\mathsf{D}}_{\textit{E}_1\textit{E}_2}\\
&=\mathbb{I}_{\textit{lab}_A}\otimes\int d\Bar{x}_1 d\Bar{x}_2\;\Delta(\Bar{x}_1,\Bar{x}_2)\ketbra{\Bar{x}_1}{\Bar{x}_1}_{\textit{E}_1}\otimes\ketbra{\Bar{x}_2}{\Bar{x}_2}_{\textit{E}_2},
\end{align}
where $\Delta(\Bar{x}_1,\Bar{x}_2):=\sqrt{(t_2-t_1)^2-(x_2-x_1)^2}$.
Accordingly, $\hat{\mathsf{D}}_{\textit{E}_1\textit{E}_2}$ provides the corresponding spacetime distance between two 
coordinate states, i.e. $\Delta(\Bar{x}_1,\Bar{x}_2)$.
Let us now transform~\eqref{distance operator}, 
using the map introduced in Eq.~\eqref{quantum-controlled Lorentz coordinate}
\begin{align}
\hat{\mathcal{S}}_{CA}\hat{\mathcal{D}}^C\hat{\mathcal{S}}^\dagger_{CA}&= \int dv\mathcal{P}_{CA}\ketbra{v}{v}_{\textit{A}} \mathcal{P}_{CA}\otimes \hat{\Lambda}_{-v}\hat{\mathsf{D}}\hat{\Lambda}_{v}\\
 &=\int dv\ketbra{-v}{-v}_{\textit{C}}\otimes\hat{\mathsf{D}}_{\textit{E}_1\textit{E}_2}\\
&=\mathbb{I}_{\textit{lab}_{\textit{C}}}\otimes\hat{\mathsf{D}}_{\textit{E}_1\textit{E}_2}=\hat{\mathcal{D}}^{A},
\end{align}
where $\hat{\Lambda}_{-v}\hat{\mathsf{D}}_{\textit{E}_1\textit{E}_2}\hat{\Lambda}_{-v}=\hat{\mathsf{D}}_{\textit{E}_1\textit{E}_2}$ stems directly from  the invariance of $\Delta(\Bar{x}_1,\Bar{x}_2)$ and $d\Bar{x}$. 
The spacetime distance operator is left untouched by the map~\eqref{quantum-controlled Lorentz coordinate}, hence each ``quantum
inertial observer'' measures the same spacetime
distance of the considered pair of events.
  
The same transformation can straightforwardly be applied to 
the events parametrized by energy-momentum pair of coordinates.
This proposal  extends the notion of Lorentz covariance
to inertial observes in a quantum superposition of 
velocities.
Besides, upon enlarging the symmetry
group of the spacetime, 
we can expect that a similar treatment, extended to 
general coordinate transformations beyond Minkowski spacetime applies.


\section{Conclusions}

The extension  of the reference frame symmetry transformations to the quantum realm has been thoroughly discussed for the case of Galilean symmetry group~\cite{QRF}. Despite important developments towards relativistic formulation of
QRFs~\cite{QRFspin,spacetimeQRF}, a formulation of Lorentz symmetry for QRFs was lacking. In this paper, we develop the notion of Lorentz covariance for QRFs and discuss new phenomena of superpositions of time dilation and length contraction that can only be explained if the reference frames are both \textit{ relativistic} and \textit{quantum mechanical}. 

We worked in a $1+1$-spacetime, where the Lorentz group reduces to the abelian group of one dimensional boosts. In section~\ref{Spacetime states and Probability}, following the proposal for a covariant 
formulation of quantum mechanics~\cite{CQM,CQM1},
we formulated quantum mechanics that treat time and space symmetrically. In the formulation, the quantum mechanical state is given independently of any notion of a preferred or spatial division of spacetime. It describes the system in arbitrary regions of spacetime and hence generalises the standard 
picture in which the quantum state is specified at a given time. It is shown that such a \textit{quantum spacetime state} transforms covariantly under the action of the Lorentz symmetry group. In contrast to Page and Wootters mechanism~\cite{P-W,HistoryDirac,HistoryScalar}, we do not introduce a clock by adding a quantum degree of freedom to the rest of the systems and dynamically constraining them.
However, as we have shown, the dynamics with respect to the second observer, A, is no longer free. It would be interesting to investigate under which conditions, if at all, both observers A and C observe (approximately) free dynamics.  
In section~\ref{RQRF_formalism} we constructed a map that switches between the descriptions given from different RQRFs. We started from an ``external view'' in which a quantum state for all systems is given. Since it is assumed that no external reference frame for the Lorentz group is given, we averaged the state with respect to the symmetry group and arrive at a perspective-neutral state. By choosing a specific RQRFs, we defined 
a unitary Lorentz transformation, controlled 
by the momentum of the RQRF, which leads to
one perspectival space. Finally, by
suitably composing the transformations, we derived the 
map that transform between all the perspectival states.

The above procedure follows the one of Refs.~\cite{QRFrelational}, but the cases considered there differ from the present one. In that work, the group averaging operator commutes with the Hamiltonian of the global system, 
i.e. the group of QRF transformations is a symmetry 
with respect to the perspectiveless dynamics. 
This is not the case if one assumes the perspectiveless dynamics to be relativistic free evolution and the averaging is taken with respect to the Lorentz group. In order to circumvent this problem, we started here already from a perspectival view of a certain RQRF, $C$, in which the dynamics of the system external to the frame are assumed to be free as given in Sec.~\ref{Spacetime states and Probability}. This could be advantageous from an experimental point of view, since all our observations are already made from a perspectival point of view, namely from the frame of our (macroscopic) laboratories. With respect to this frame, the dynamics of the (non-interacting) relativistic particles is free. 

In section~\ref{RQRF_formalism}, we explore the phenomenological consequences resulting 
from moving to the description of a reference frame in a superposition of 
momenta. In particular, we analyse  quantum superposition of genuine special-relativistic effects, such as time dilations, length contractions and the invariance of spatiotemporal distance between two events. In its original classical relativistic context, these effects resulted from Einstein's operational approach to spacetime, which is based on how clock's ticking rates and rod lengths transforms between inertial observers. Our work can be seen as extending this operational approach to the quantum domain.

The present approach has  limits of applicability. This limitation, however, is purely inherited from the known difficulties in formulating  relativistic 
(single-particle) quantum mechanics, and is known to be overcame by extending the formalism to quantum fields: Initially localized particles on compact supports can propagate superluminally enabling signalling between spacelike separated agents. Nevertheless, relativistic causality is restored by taking a suitable limit, so that we can provide a physically testable  scenario.

Our work can be placed in a broader research program
aimed to analyse a (semiclassical) regime of a quantum 
spacetime~\cite{QRFmassiveobjects,QRFmassesinsuperposition,spacetimeQRF,QRFspacetime}. From this perspective it is important to extend the notion of general covariance to QRFs, i.e. to apply the full diffeomorphism symmetry group to QRFs. Although here we worked only with 
Lorentz boosts in a fixed background, they constitute a distinctive subgroup of diffeomorphisms. A natural continuation of our work would be extending the notion of covariance for the entire Poincaré group for QRFs.

\vspace{0.25cm}

\section*{Acknowledgements}

L.A. and {\v C}.B. acknowledge financial support by
the Austrian Science Fund (FWF) through BeyondC
(F7103-N48). E.C.-R. is supported by an ETH Zurich Postdoctoral Fellowship and acknowledges financial support from the Swiss National Science Foundation (SNSF) via the National Centers of Competence in Research QSIT and SwissMAP, as well as the project No. 200021 188541. This publication was made possible
through the support of the ID 61466 and ID 62312
grants from the John Templeton Foundation, as part
of The Quantum Information Structure of Spacetime
(QISS) Project (qiss.fr). The opinions expressed in
this publication are those of the authors and do not
necessarily reflect the views of the John Templeton
Foundation.

\bibliographystyle{unsrtnat}
\bibliography{References}
\appendix
\section{Superposition of the wave packet extensions}\label{alternative_lenght}

In this appendix we describe a wave packet with superposed extensions as a result of the superposition of Lorentz length contractions.
 
\begin{figure*}[ht]
\centering
\includegraphics[width=380pt]{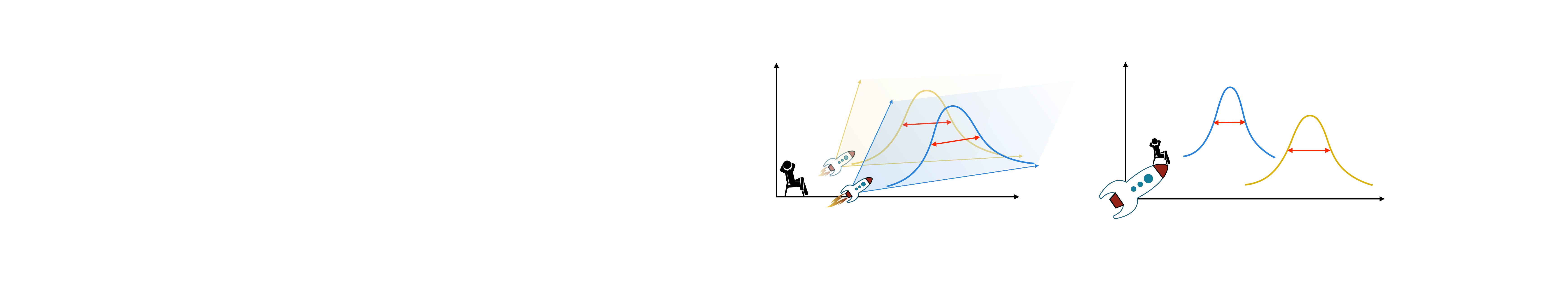}
\caption{\textit{Superposition of Gaussian states with special-relativistic contracted widths}:  (Left) A superposition of two identical Gaussian states lying on two tilted hypersurfaces from the point of view of the observer ($C$) at the ground. (Right) For an observer who moves with the space ship $A$ the superposition of Gaussian states lies on two simultaneity surfaces.
The special-relativistic length contraction is witnessed by the contracted widths of the Gaussians.
}
\label{sup_lengthcontraction}
\end{figure*}

Consider the situation where, from $C$'s perspective, one
prepares a superposition of
two Gaussian wavepackets $f_1$ and $f_2$, each of which lies on a different 
simultaneity surface, so that the two surfaces are tilted with respect to each other (see Fig.~\ref{sup_lengthcontraction}).
Each simultaneity surface is geometrically defined by 
$t_i(x)=\alpha_ix$, for $i=1,2$. The gradients $\alpha_i$ are given by 
the hyperbolic rotations, $\alpha_{i}=-\tanh{\omega_{i}}$, 
where $\omega_{i}$ are the angles (rapidity) of the
hyperbolic rotations.

The joint state of system $B$ and RQRF $A$
from $C$'s perspective is given by
\begin{equation}
    \ket{\psi}^{(C)}=\ket{f_1^B}\otimes\ket{\phi_1^A}+\ket{f_2^B}\otimes\ket{\phi_1^A},
\end{equation}
where the two Gaussian states, defined on the tilted hypersurfaces, have the following form
\begin{equation*}
    \ket{f_{i}^B}=\frac{1}{\sqrt{2\pi}\sigma}\int dtdx\ket{t,x}\delta(t-\alpha_{i}x)e^{-x^2/4\sigma^2}, \mbox{ } i=1,2
\end{equation*}
and $\ket{\phi_{i}^A}$ are sharply peaked around $\omega_{i}$. The variance $\sigma^2$ characterises the width of the wave packet. 

To change to $A$'s perspective we apply the map $\hat{\mathcal{S}}_{C\rightarrow A}$ and obtain 
\begin{align}\label{sup_Contr}
\ket{\psi}^{(A)}:&=\hat{\mathcal{S}}_{C\rightarrow A}\ket{\psi}^{(C)}\nonumber\\
&=\ket{f^B_{1\omega_1}}\otimes\ket{\phi^C_1}+\ket{f^B_{2\omega_2}}\otimes\ket{\phi^C_2}.
\end{align}
Here
\begin{align*}
&\ket{f^B_{{i\omega_{i}}}}=\frac{1}{\sqrt{2\pi}\sigma}\int
dtdx\ket{\Lambda_{-\omega_{i}}(t,x)}\delta(t+t_{i})e^{-\frac{x^2}{4\sigma^2}}\\
&=\frac{1}{\sqrt{2\pi}\sigma}\int
dtdx\ket{t,x}\delta(\Lambda^0_{\omega_{i}}(t,x)+t_{i}(\Lambda^1_{\omega_{i}}(t,x)))\\
&\quad e^{-\frac{\Lambda^1_{\omega_{i}}(t,x)^2}{4\sigma^2}},
\end{align*}
and the delta function can be simplified as follows
\begin{align*}
\delta(t\cosh{\omega}(1-\tanh{\omega}^2))=\delta(t\cosh^{-1}{\omega})=|\cosh{\omega}|\delta(t),
\end{align*}
which results in 
\begin{align*}
    \ket{f^B_{{i\omega_{i}}}}&=\frac{1}{\sqrt{2\pi}\left(\frac{\sigma}{\cosh{\omega_{i}}}\right)}\int dx\ket{0,x}e^{-\frac{x^2}{4\left(\frac{\sigma}{\cosh{\omega_{i}}}\right)^2}}\\
    &=\frac{1}{\sqrt{2\pi}\sigma_{i}}\int dx\ket{0,x}e^{-\frac{x^2}{4\sigma_{i}^2}}.
\end{align*}
Finally, one notices that, in $A$'s perspective, the Gaussian states
lie on a single simultaneity surface, corresponding
to $t=0$. Second, the 
widths of the Gaussians in the two branches of the transformed state are given by
\begin{equation*}
\sigma_{i}=\frac{\sigma}{\cosh{\omega_{i}}}=\sqrt{1-\tanh{\omega_{i}}}\sigma=\frac{\sigma}{\gamma(\omega_{i})},
\end{equation*}
with $i=1,2$, 
i.e. they are contracted by
an amount of  $\gamma(\omega_{i})^{-1}<1$. 
We conclude that the state with a definite wave-packet width in $C$'s reference frame, 
transforms into a superposition of states each with Lorentz contracted wave packet width in $A$'s RQRF.
In that sense state~\eqref{sup_Contr} is an example of a quantum 
superposition of special-relativistic space contractions (see Fig.~\ref{sup_lengthcontraction} for an illustration of the state).

\subsubsection{Probing superposition in the non-relativistic regime}

We next consider a single Gaussian wave packet in space. We want to think of this wave packet as describing the amplitudes for position measurements on a quantum system in space. However, it is well known that the position operator in relativistic quantum mechanics is not well defined~\cite{Malament,Busch,Sorkin,Wightman}. For this reason, we consider the non-relativistic limit of our transformations, and explore  relativistic corrections up to the second order in $\omega_i$ to the position measurements.
Furthermore, 
we assume that the Gaussian state is prepared within a finite
time interval, which for the sake of simplicity is again described by a Gaussian distribution.

Relative to $A$'s perspective, the
state of $B$ and the 
reference frame $C$ is given by 
\begin{equation}
    \ket{\psi}^{(C)}=\ket{f^B}\otimes\ket{\phi^A},
\end{equation}
where  
\begin{align}
    \ket{f^B}&=\frac{1}{2\pi\sigma_x^2\sigma_t^2}\int dxdt \ket{t,x}e^{-(x-x_0)^2/4\sigma_x^2}e^{-(t-t_0)^2/4\sigma_t^2}\\
    &=\mathcal{C}\int dxdt \ket{t,x}e^{-(x-x_0)^2/4\sigma_x^2}e^{-(t-t_0)^2/4\sigma_t^2},
\end{align}
where $\mathcal{C}$ is a normalisation constant, and $\sigma_x$ and $\sigma_t$ are the standard deviations describing spatial and temporal extensions of the wave packet, respectively. As usual the state reference frame $A$ is taken to be in a superposition 
of velocities as in Eq.~\eqref{sharp_momenta}. 

Now we adopt $A$'s point of view by means of the map $\hat{\mathcal{S}}_{C\rightarrow A}$, 
obtaining
\begin{align*}
    \ket{\psi}^{(A)}:&=\hat{\mathcal{S}}_{C\rightarrow A}\ket{\psi}^{(C)}\\
    &=\ket{f^B_{-\omega_1}}\otimes\ket{\phi^C_1}+\ket{f^B_{-\omega_2}}\otimes\ket{\phi^C_2}.
\end{align*}
where
\begin{align}\label{gaussian}
    &\ket{f^B_{-\omega_i}}=\int dxdt \ket{\Lambda_{-\omega_i}(t,x)}e^{-(x-x_0)^2/4\sigma_x^2}e^{-(t-t_0)^2/4\sigma_t^2} \nonumber\\ 
    &=\int dt dx \ket{t,x}e^{-(\Lambda^1_{\omega_i}(t,x)-x_0)^2/4\sigma_x^2} e^{-(\Lambda^0_{\omega_i}(t,x)-t_0)^2/4\sigma_t^2}.
\end{align}
We next take the non-relativistic limit of Eq.~\eqref{gaussian}, that is, we assume  
$\vert \omega_i \vert \ll 1$ for $i= 1, \ 2$.
This justifies the following expansion 
\begin{equation*}
    \Lambda_\omega=
    \begin{pmatrix}
    \cosh(\omega_i)&-\sinh(\omega_i)\\
    -\sinh(\omega_i)&\cosh(\omega_i)
    \end{pmatrix}\approx
    \begin{pmatrix}
    1+\omega_i^2/2& -\omega_i\\
    -\omega_i&1+\omega_i^2/2
    \end{pmatrix}.
\end{equation*}
Finally, the state in Eq.~\eqref{gaussian} assumes the form:
\begin{align*}
    \ket{f^B_{-\omega_i}}&=\int dt dx \ket{t,x}e^\frac{-((1+\frac{\omega_i^2}{2})x-\omega_i t-x_0)^2}{4\sigma_x^2}e^\frac{-((1+\frac{\omega_i^2}{2})t-\omega_i x-t_0)^2}{4\sigma_t^2}\\
    &=\int dt dx \ket{t,x}\phi^{t_0,x_0}_{\omega_i}(t,x)
\end{align*}
where the $\ket{t,x}$ has here the following form
\begin{equation}\label{non-rel_state}
    \ket{t,x}=\int dp e^{i\frac{p^2}{2m}t-ipx}\ket{p}.
\end{equation}

We will next measure the spacetime location  of particle $B$ conditional a postselected measurement result on $C$. For the measurement of $B$ we will consider the
POVM element $T^B_{(t',x')}:=\ketbra{t',x'}{t',x'}$, with the non-relativistic spacetime kets, as given in Eq.~\eqref{non-rel_state}. 
We suppose that the QRFs $C$ is measured in the bases spanned by  $\ket{\pm}=\ket{\phi_1}\pm\ket{\phi_2}$. With a suitable choice of parameters we can approximate that $\braket{\phi_1|\phi_2}\approx 0$. Hence, depending on the postelected result on $C$, we are left with the following \sout{marginal} conditional state of system $B$:
\begin{align*}
    \rho_B&=\text{Tr}_C[(\ketbra{\pm}{\pm}\otimes I_B)\ketbra{\psi}{\psi}^{(A)}]\\
    &=\braket{\phi_1|\phi_1}\ketbra{f^B_{-\omega_1}}{f^B_{-\omega_1}}+\braket{\phi_2|\phi_2}\ketbra{f^B_{-\omega_2}}{f^B_{-\omega_2}}\\
    &\pm\braket{\phi_1|\phi_1}\braket{\phi_2|\phi_2}(\ketbra{f^B_{-\omega_1}}{f^B_{-\omega_2}}
    +\ketbra{f^B_{-\omega_2}}{f^B_{-\omega_1}}).
\end{align*}
The probability for detecting $B$ at spacetime location $(t',x')$ is given by 
\begin{align*}
 p_{(t',x')}:=&\braket{\phi_1|\phi_1}|\braket{f^B_{-\omega_1}|t',x'}|^2+\braket{\phi_2|\phi_2}|\braket{f^B_{-\omega_2}|t',x'}|^2\\
 &\pm 2\braket{\phi_1|\phi_1}\braket{\phi_2|\phi_2}\text{Re}[\braket{f^B_{-\omega_1}|t',x'}\braket{t',x'|f^B_{-\omega_2}}],
\end{align*}
where 
\begin{align*}
   &\braket{f^B_{-\omega_{1/2}}|t',x'}=\int dt dx \phi^{t_0,x_0}_{\omega_{1/2}}(t,x)\braket{t,x}{t',x'}\\
   &=\int dt dx \phi^{t_0,x_0}_{\omega_{1/2}}(t,x)\left(\frac{m}{i(t- t')}\right)^{\frac{1}{2}}e^{i\frac{m}{2(t-t')}(x-x')^2}
\end{align*}
and $\braket{t',x'|t,x}=\bra{x'}\hat{U}(\Delta t)\ket{x}=\left(\frac{m}{i(\Delta t)}\right)^{\frac{1}{2}}e^{i\frac{m}{2(\Delta t)}(\Delta x)^2}$ represents
the Schr\"{o}dinger propagator.
As a result, 
we observe that from perspective $A$,
the probability of detecting system $B$
 in $(t',x')$ has a contribution 
corresponding to the interference 
between $\ket{f^B_{-\omega_1}}$ and  
$\ket{f^B_{-\omega_2}}$.
We can therefore observe the coherences of a ``quantum superposition of length contraction'' by post-selecting on $C$ being either in state $\ket{+}$ or $\ket{-}$ and probing the spacetime probability distribution of $B$ in the non-relativistic limit. 

\section{Perspectival states}\label{B}
We show the intermediate steps in the derivation of perspectival states from Eq.~\eqref{perspective-less} to Eq.~\eqref{psiam}.
\begin{widetext}
\begin{align}\label{derivation_A-persp}
\ket{\psi}^{(A)}=\hat{\V}_A\int d\omega
\ket{\psi_\omega}_{ABC}
&=\hat{\V}_A\int d\alpha d\beta d\gamma \ket{\Lambda^1_{\omega+\alpha} k_A}\ket{\Lambda^1_{\omega+\beta} k_B}\ket{\Lambda^1_{\omega+\gamma} k_C}\psi(\alpha,\beta,\gamma)\\
&=\int d\alpha \int d\lambda\int d\omega\ket{\Lambda^1_\lambda k_A}\braket{\Lambda^1_\lambda k_A|\Lambda^1_{\omega+\alpha} k_A}\otimes {U^{BC}}(\Lambda_{-\lambda})\ket{\phi_\omega(\Lambda_\alpha k_A)}_{BC}\\
&=\int d\alpha\int d\omega \ket{\Lambda^1_{\alpha+\omega}k_A}\otimes\ket{\phi_{-\alpha}(\Lambda_\alpha k_A)}_{BC}\\
&=\int d\omega\ket{\Lambda^1_\omega k_A}\otimes\int d\alpha\ket{\phi_{-\alpha}(\Lambda_\alpha k_A)}_{BC}\\
&=\ket{\Omega}\otimes\ket{\psi}_{BC},
\end{align} 
\end{widetext}
where $\tilde{f}_\alpha(\Lambda_\beta k_A):=\int dtdx e^{i\Lambda^0_\beta k_At-i\Lambda^1_\beta k_Ax}f_\alpha(t,x)$ and
$\ket{\phi_\omega(\Lambda_\alpha k_A)}_{BC}:=\int d\beta d\gamma\ket{\Lambda^1_{\omega+\beta} k_B}\ket{\Lambda^1_{\omega+\gamma} k_C}\psi(\alpha,\beta,\gamma)$.


\section{Superposition of boosts}\label{superposition of boosts}
In this appendix we give the full 
derivation of the state  in 
Eq.~\eqref{karta}, under  the
conditions expressed in Eqs.~\eqref{sup_boosts.psi_c} and \eqref{sharp_momenta}. We have
\begin{widetext}
\begin{align}
    &\ket{\psi}_{BC}:=\hat{\mathcal{S}}_{C\rightarrow A}\ket{\psi}_{AB}=\int d\alpha  \ket{\Lambda^1_{-\alpha} k_C}\braket{\Lambda^1_\alpha k_A|f^A_\alpha}\otimes U_B(\Lambda_{-\alpha})\ket{f^B(t_B,x_B)}\\
    &=\int d\alpha d\alpha'\int dt_Adx_A\ket{\Lambda^1_{-\alpha} k_C}\underbrace{\braket{\Lambda^1_\alpha k_A|\Lambda_{\alpha'}k_A}}_{\delta(\alpha-\alpha')}e^{i\Lambda^0_{\alpha'}k_At+i\Lambda^1_{\alpha'}k_Ax_A}f_A(t,x_A)\otimes\ket{f^B_{-\alpha}}\\
    &=\int d\alpha\int dt_A\ket{\Lambda^1_{-\alpha} k_C}e^{i\Lambda^0_{\alpha}k_At}g_A(t)(\delta(\alpha-\omega_1)+\delta(\alpha-\omega_2))\otimes\ket{f^B_{-\alpha}}\label{3}\\
   &=\tilde{g}_C(\Lambda^0_{\omega_1}k_C)\ket{\Lambda^1_{-\omega_1}k_C}\otimes\ket{f^B_{-\omega_1}}+\tilde{g}_C(\Lambda^0_{\omega_2}k_C)\ket{\Lambda^1_{-\omega_2}k_C}\otimes\ket{f^B_{-\omega_2}}\label{gtilde}\\
    &=\int dt_C\underbrace{\frac{m_C}{m_A}g_A(\frac{m_C}{m_A}t_C)}_{:=g_C(t_C)}\left(\ket{\Lambda^1_{-\omega_1}k_C}e^{i\Lambda^0_{\omega_1}k_Ct_C}\otimes\ket{f^B_{-\omega_1}}+\ket{\Lambda^1_{-\omega_2}k_C}e^{i\Lambda^0_{\omega_2}k_Ct_C}\otimes\ket{f^B_{-\omega_2}}\right)\\
    &=\int dt_Cdx_C\int d\alpha\ket{\Lambda^1_{-\alpha}k_C}e^{i\Lambda^0_{\alpha}k_Ct_C+i\Lambda^1_{\alpha}k_Cx_C}
    (\underbrace{g_C(t_C)\phi^{1}_C(x_C)}_{:=f^C_1(t_C,x_C)}\ket{f^B_{-\omega_1}}+\underbrace{g_C(t_C)\phi^{2}_C(x_C)}_{:=f^C_2(t_C,x_C)}\ket{f^B_{-\omega_2}})\label{6}\\
    &=\int dt_Cdx_C\int d\alpha\ket{\Lambda_{-\alpha}k_C}e^{i\Lambda^0_{\alpha}k_Ct_C+i\Lambda^1_{\alpha}k_Cx_C}
    (f^C_1(t_C,x_C)\ket{f^B_{-\omega_1}}+f^C_2(t_C,x_C)\ket{f^B_{-\omega_2}}),
\end{align}
\end{widetext}
 where we defined $\tilde{g}_C(\Lambda^0_{\omega_1}k_C):=\int t_A e^{i\Lambda^0_{\alpha}k_At}g_A(t)$,
$k_A=k_C\frac{m_A}{m_C}$ and $t_C=\frac{m_A}{m_C}t_A$. We also identified the spacetime function of RQRF $C$ as $f^C_i(t_C,x_C):=g_C(t_C)\phi^C_i(x_C)$, with $\phi^C_i(x_C)$
 such that its Fourier transform is a 
delta function centered in $\omega_i$, for $i=1,2$.

\section{Superposition of simultaneity surfaces}\label{superposition of simultaneity surfaces}
In this appendix we give the detailed derivation 
of Eq.~\eqref{psibc_supsli1}, 
where the states for $A$ and $B$ are 
given in Eqs.~\eqref{slice_b} and
\eqref{sharpsupt_A}. We obtain 
\begin{widetext}
\begin{align*}
&\ket{\psi}_{BC}:=\hat{\mathcal{S}}_{C\rightarrow A}\ket{\psi}_{AB}
=\int d\alpha\ket{\Lambda^1_{-\alpha}k_C}\braket{\Lambda^1_{\alpha}k_A|f^A_\alpha}\otimes U^B(\Lambda_{-\alpha})\ket{f^B}\\
&=\int d\alpha\int d\alpha'\int dt dx_A\ket{\Lambda^1_{-\alpha}k_C}\braket{\Lambda^1_{\alpha}k_A|\Lambda^1_{\alpha'}k_A}e^{i\Lambda^0_{\alpha'}k_At-i\Lambda^1_{\alpha'}k_Ax_A}f^A(t
,x_A)\otimes\ket{f^B_{-\alpha}}\\
&=\int d\alpha\int dt dx_A\ket{\Lambda^1_{-\alpha}k_C}e^{i\Lambda^0_{\alpha}k_At-i\Lambda^1_{\alpha}k_Ax_A}\delta(t-t_A)\phi^A(x_A)\otimes\ket{f^B_{-\alpha}}\\
&=\int d\alpha\int dt\ket{\Lambda^1_{-\alpha}k_C}e^{i\Lambda^0_{\alpha}k_At}\delta(t-t_A)(\delta(\alpha-\omega_1)+\delta(\alpha-\omega_2))\otimes\ket{f^B_{-\alpha}}\\
&=\int dt\left(\ket{\Lambda^1_{-\omega_1}k_C}e^{i\Lambda^0_{\omega_1}k_At}\delta(t-t_A)\otimes\ket{f^B_{-\omega_1}}+\ket{\Lambda^1_{-\omega_2}k_C}e^{i\Lambda^0_{\omega_2}k_At}\delta(t-t_A)\otimes\ket{f^B_{-\omega_2}}\right)\\
&=\int dt\int dx_C\int d\alpha\delta(t-t_A)e^{i\Lambda^0_{\alpha}k_At-i\Lambda^1_\alpha k_Cx_C}\left(\ket{\Lambda^1_{-\alpha}k_C}\phi^C_1(x_C)\otimes\ket{f^B_{-\omega_1}}+\ket{\Lambda^1_{-\alpha}k_C}\phi^C_2(x_C)\otimes\ket{f^B_{-\omega_2}}\right)\\
&=\int dt\int dx_C\int d\alpha\delta(t_C-t'_C)e^{i\Lambda^0_{\alpha}k_Ct_C-i\Lambda^1_{\alpha} k_Cx_C}\left(\ket{\Lambda^1_{-\alpha}k_C}\phi^C_1(x_C)\otimes\ket{f^B_{-\omega_1}}+\ket{\Lambda^1_{-\alpha}k_C}\phi^C_2(x_C)\otimes\ket{f^B_{-\omega_2}}\right)\\
&=\int dt\int dx_C\int d\alpha\left(\ket{\Lambda^1_{-\alpha}k_C}e^{i\Lambda^0_{\alpha}k_Ct_C-i\Lambda^1_{\alpha} k_Cx_C}f^C_1(t_C,x_C)\otimes\ket{f^B_{-\omega_1}}+\ket{\Lambda^1_{-\alpha}k_C}e^{i\Lambda^0_{\alpha}k_Ct_C-i\Lambda^1_\alpha k_Cx_C}f^C_2(t_C,x_C)\otimes\ket{f^B_{-\omega_2}}\right)\\
&=\ket{\Lambda^1_{-\omega_1}k_C}e^{i\Lambda^0_{\omega_1}k_Ct_C}\otimes\ket{f^B_{-\omega_1}}+\ket{\Lambda^1_{-\omega_2}k_C}e^{i\Lambda^0_{\omega_2}k_Ct_C}\otimes\ket{f^B_{-\omega_2}},
\end{align*}
\end{widetext}
where we have defined $t'_C:=\frac{m_A}{m_C}t_A$, and $t_C:=\frac{m_A}{m_C}t$ which
determine either a time dilation or a contraction, depending
on the ratio of  masses. Furthermore we denoted
the spacetime functions $f^C_{i}(t_C,x_C):= \delta(t_C-t'_C)\phi^C_i(x_C)$, which 
describes a relativistic particle with momentum $\Lambda_{\omega_{i}}k_C$, located in the simultaneity surface labelled by $t_C$.

\end{document}